\documentclass[a4paper,11pt]{article}
\usepackage{jcappub} % for details on the use of the package, please see the JINST-author-manual
\usepackage{amsmath,amsfonts,amssymb,mathtools,bm,cancel}
\usepackage{comment}
\usepackage{graphicx}
\usepackage{hyperref,cleveref}
\usepackage[numbers]{natbib}
\usepackage[T1]{fontenc} 
\usepackage{booktabs}
\usepackage{multirow}
\usepackage[dvipsnames]{xcolor}

\newcommand{\de}{\mathrm{d}}

\newcommand{\fnl}{f_{\rm NL}}

\newcommand{\Mk}{{\mathcal M}}

\newcommand{\Zker}{{\mathcal Z}}
\newcommand{\bfi}{b_{\phi}}
\newcommand{\bfd}{b_{\phi\delta}}

\newcommand{\be}{\begin{equation}}
\newcommand{\ee}{\end{equation}}
\newcommand{\bea}{\begin{eqnarray}}
\newcommand{\eea}{\end{eqnarray}}
\newcommand{\bdm}{\begin{displaymath}}
\newcommand{\edm}{\end{displaymath}}

\newcommand{\avg}[1]{\ensuremath{\left\langle \,#1\, \right\rangle}}
\newcommand{\xv}{\mathbf{x}}

\newcommand{\kv}{\mathbf{k}}

\newcommand{\qv}{\mathbf{q}}

\title{\boldmath Bispectrum constraints on Primordial non-Gaussianities with the eBOSS DR16 quasars}

\author[a,1]{Marina S. Cagliari,\note{Corresponding author}}
\author[b,c,d]{Matilde Barberi-Squarotti,}
\author[e]{Kevin Pardede,}
\author[b,d]{Emanuele Castorina,}
\author[f,e]{Guido D'Amico}

\affiliation[a]{Laboratoire d’Annecy-le-Vieux de Physique Theorique (LAPTh),
CNRS/USMB, \\ 9 Chemin de Bellevue BP110, F-74941 Annecy, France}
\affiliation[b]{Dipartimento di Fisica ‘Aldo Pontremoli’, Università degli Studi di Milano, \\ Via Celoria 16, I-20133 Milan, Italy}
\affiliation[c]{INAF, Osservatorio Astrofisico di Brera-Merate, Via Brera 28, 20121, Milan, Italy}
\affiliation[d]{INFN, Sezione di Milano, Via Celoria 16, 20133 Milan, Italy}
\affiliation[e]{INFN, Gruppo Collegato di Parma, Parco Area delle Scienze 7/A, I-43124 Parma, Italy}
\affiliation[f]{Dipartimento di Scienze Matematiche, Fisiche e Informatiche, Università di Parma, Viale delle Scienze 7/A, I-43124 Parma, Italy}

\emailAdd{marina.cagliari@lapth.cnrs.fr}
\emailAdd{matilde.barberi@unimi.it}
\emailAdd{kevinfranklysamuel.pardede@unipr.it}
\emailAdd{emanuele.castorina@unimi.it}
\emailAdd{guido.damico@unipr.it}

\abstract{We present constraints on $\fnl$, the parameter quantifying the amplitude of local Primordial Non-Gaussianities (PNG), from a combined analysis of the tree-level power spectrum and bispectrum of Data Release $16$ (DR16) of the extended Baryon Oscillation Spectroscopic Survey (eBOSS) quasar sample. In our analysis, we use the power spectrum measured with the optimal redshift weights that maximize the local PNG information together with the bispectrum estimated with the standard Feldman-Kaiser-Peacock weights. In the modeling, we incorporate the global and radial integral constraint corrections both in the power spectrum and in the bispectrum, for which we observe that only the radial integral constraint correction has a significant impact. Our constraints read $-6 < \fnl < 20$ at $68\%$ confidence level and improve by $\sim 16\%$ over the previous power spectrum analysis of the same dataset. We observe the same improvement over the power spectrum analysis when the quasar response to PNG is lower. In this case, we find $-23 < \fnl < 14$ at $68\%$ confidence level. Our findings are consistent with the Fisher matrix expectations.}


\begin{document}
\maketitle
\flushbottom

\section{Introduction and main results}\label{sec:intro}

The Large Scale Structure (LSS) distribution of the Universe that we observe today results from the gravitational evolution of the primordial curvature fluctuations. The late-time LSS thus encodes information about the very early Universe, that we can probe with $n$-point statistics of galaxy samples. In particular, the galaxy distribution could test for the presence of Primordial Non-Gaussianities (PNG), possibly produced during Inflation \citep[for a review see][]{Baumann:2009ds}, allowing us to distinguish between different inflationary scenarios. However, we expect PNG to be subdominant with respect to the Gaussian component of the perturbations, making their detection challenging.

In this work, we focus on PNG of the so-called local type. Local PNG are parametrized by a single number, $\fnl$, which controls the amount of non-linearity in the primordial gravitational potential, $\Phi_P (\bf{x})$, via $\Phi_P = \varphi + f_{\rm NL} (\varphi^2-\avg{\varphi^2})$, where $\varphi$ is a Gaussian random field. In the case of single-field inflation models local PNG are exactly zero \citep{Maldacena:2002vr,Creminelli:2004yq,Cabass:2016cgp}; on the other hand, multi-field inflation models predict $\fnl \sim \mathcal{O}(1)$ \citep{Senatore:2010wk,Alvarez:2014}, and could be constrained by a strong experimental bound on $\fnl$. Currently, the tightest constraints on local PNG come from the measurements of the bispectrum, i.e. the harmonic transform of the $3$-point function, of the Cosmic Microwave Background (CMB) anisotropies measured by the \textit{Planck} satellite and read $\fnl = 0.8 \pm 5$ \cite{2020A&A...641A...9P}. We expect upcoming CMB experiments to tighten these bounds by no more than $50\%$ \citep{CMB-S4:2016ple}. 

An alternative to CMB measurements, which are intrinsically two-dimensional maps, is to use the LSS of the Universe, which is three-dimensional and could thus probe a larger fraction of the primordial fluctuations. Moreover, and contrary to the CMB case, PNG start to manifest in the LSS at the two-point function level. This is due to the coupling between large and small scales introduced by the initial non-linearity in the gravitational potential $\Phi_P$, which, as first realized by \cite{Dalal:2008}, modifies the relation between the galaxy density field $\delta_g$ and the underlying matter perturbations. Specifically, this relation, known as galaxy bias, will now contain terms proportional to the amplitude of PNG, of the form $\delta_g(\mathbf{x},z) \supset (f_{\rm NL} b_\phi(z) \Phi_p (\mathbf{x}) +  f_{\rm NL} b_{\phi \delta}(z)  \delta_m(\mathbf{x},z) \Phi_p (\mathbf{x}) + ...)$.

This result spurred enormous interest in constraining local PNG with galaxy surveys, as any statistic of the galaxy density field, including, as mentioned above, the $2$-point ones, will be sensitive to $f_{\rm NL}$. The price to pay is that the signal one is trying to measure is highly degenerate with the largely unknown value of the new bias parameters $b_\phi$ and $b_{\phi \delta}$. Any prior assumption about the latter would for instance imply that different tracers cannot be combined together or with the CMB constraints.\footnote{See for example \cite{Barreira:2020kvh,2022JCAP...01..033B}.}

The most stringent LSS-based constraints on $\fnl$ came recently from the power spectrum analysis of the Dark Energy Spectroscopic Instrument \citep[DESI;][]{desi}, where, combining the quasar (QSO) and luminous red galaxy (LRG) samples, an uncertainty of $\sigma_{\fnl} \sim 9$ was reported \citep{DESI:2024oco}. This bound significantly improved over previous power spectrum constraints, for the most part based on the QSO catalogs collected by the Sloan Digital Sky Survey (SDSS) \citep{Castorina2019,Mueller:2022dgf,Cagliari:2023mkq}. 
The methodology for a PNG analysis of the power spectrum is very mature by now and often built on the optimal quadratic estimator discussed in ref.~\citep{Castorina2019}. Future improvements in such bounds on $f_{\rm NL}$ will likely come from larger and denser galaxy samples.\footnote{Note that the optimal methods introduced in ref.~\cite{Castorina2019} do not perform a full inverse covariance weighting of the data \cite{1997ApJ...480...22T}. }

A way to further improve the power spectrum-based constraints is to analyze the power spectrum in combination with the bispectrum \citep{2010AdAst2010E..73L,Sefusatti:2007ih,MoradinezhadDizgah:2020whw,DAmico:2022gki,Cabass:2022ymb}. However, very low signal-to-noise ratios plague the bispectrum measurements even more than the power spectrum one, making the analysis extremely challenging.
In this work, we run a power spectrum plus bispectrum analysis of the extended Baryon
Oscillation Spectroscopic Survey Data Release 16 Quasar catalog (eBOSS QSO DR16), the precursor of DESI. Our goal is to take special care in understanding the best way to maximize the information extraction from the data given our current tools, such as the optimal power spectrum estimator for $\fnl$, and trying to both improve the modeling of observational effects in the bispectrum, estimating the limits of possible approximations. 
This is relevant given that current and upcoming surveys such as DESI \citep{desi}, \textit{Euclid} \citep{euclid}, SPHEREx \citep{spherex}, and the Vera C. Rubin Legacy Survey of Space and Time \citep{lsst}, could reach the theoretical threshold of $\sigma(f_{\rm NL}) \sim 1$ with a combination of power spectrum and bispectrum data \cite{Karagiannis:2018jdt,CosmicVisions21cm:2018rfq,Sailer:2021yzm,Cabass:2022epm,Braganca:2023pcp,MoradinezhadDizgah:2018lac,2024PhRvD.109l3511H}.

We analyze the combination of the optimally weighted power spectrum with the bispectrum and compare these results with either the optimal power spectrum-only or the traditional power spectrum measurements, based on the widely used, but sub-optimal, Feldman-Kaiser-Peacock \cite{FKP} estimator, combined with the bispectrum. In our analysis, we assume the value of the bias coefficients $b_\phi$ and $b_{\phi \delta}$ is known, and can be parametrized by a single number $p = 1.0, \, 1.6$, for QSO respectively with a higher or lower response to the presence of local PNG. Our tightest bounds correspond to the combination of the optimal power spectrum with the bispectrum and they read
\begin{equation}
\begin{cases}
    -6 < \fnl < 20 \, ,  \quad  \,\, \,68\%\,\text{c.l.} \, ,\\
    -18 < \fnl < 33 \, ,  \quad  \, 95\%\,\text{c.l.} \, ,
\end{cases}  \text{for } p=1.0\,,
\label{eq:best-bounds_p1p0}
\end{equation}
and
\begin{equation}
\begin{cases}
    -23 < \fnl < 14 \, ,  \quad 68\%\,\text{c.l.}  \, , \\
    -40 < \fnl < 33 \, ,  \quad 95\%\,\text{c.l.} \, ,
\end{cases}\text{for } p=1.6\, .
\label{eq:best-bounds_p1p6}
\end{equation}
Compared to the optimal power spectrum bounds of ref.~\citep{Cagliari:2023mkq}, adding the bispectrum information reduces the errorbar on $\fnl$ by $\sim16\%$ for both values of $p$. On the other hand, when comparing our best results with the combination of the FKP weighted power spectrum with the bispectrum the improvement is of the order of $19\%$ for $p=1.0$ and $37\%$ for $p=1.6$, which shows the importance of using the optimal weighting scheme for the power spectrum measurement even when combined with the bispectrum. To further emphasize this last point, we note that the constraints from the standard power spectrum plus bispectrum analysis are less stringent than the bounds from the optimal power spectrum analysis. Additionally, we tested a Singular Value Decomposition (SVD) compression algorithm for the bispectrum to check the covariance matrix and possibly improve the signal-to-noise \citep{Philcox:2020zyp}. We combined this compressed bispectrum with the power spectrum. However, we did not observe any improvement in this configuration.

The paper is organized as follows: in section~\ref{sec:data} we briefly present the eBOSS DR16Q sample, the measurements of the power spectrum and bispectrum, and the bispectrum compression procedure; in section~\ref{sec:analysis} we discuss the tree-level power spectrum and bispectrum models, how we perform their convolution with the window function and correct for integral constraint (IC) effects; in section~\ref{sec:preanalysis} we first present the Fisher information and cosine analyses and then we discuss our constraints on $\fnl$ in section~\ref{sec:results} we conclude and summarize our results in section~\ref{sec:conclusions}.

All the codes, scripts, measurements, and MonteCarlo Markov chains used in this work are freely accessible at \url{https://github.com/mcagliari/eBOSS-DR16-QSO-OQE/tree/PplusB}.

\section{Data} \label{sec:data}
\subsection{The eBOSS QSO Sample: data and mocks}\label{sec:data-mocks}

In this study, we use the eBOSS DR16Q sample \citep{2020MNRAS.498.2354R,2020ApJS..250....8L}, which is part of the Sloan Digital Sky Survey IV (SDSS-IV) program \citep{2017AJ....154...28B}. The DR16Q sample includes $343\,708$ quasars within the redshifts $[0.8, 2.2]$. It is divided into two fields: the North Galactic Cap (NGC), covering $2,924 \, \text{deg}^2$, and the South Galactic Cap (SGC), covering $1,884 \, \text{deg}^2$. The NGC and SGC samples respectively have an effective volume of $13.62 \, (\text{Gpc}/h)^3$ and $8.76 \, (\text{Gpc}/h)^3$ \citep{Chudaykin:2022nru}. The NGC field has a mean number density of $n \approx 1.8 \times 10^{-5} \, (\text{Mpc}/h)^{-3}$, while the SGC field has a slightly lower density of $n \approx 1.6 \times 10^{-5} \, (\text{Mpc}/h)^{-3}$, its density being approximately $10\%$ lower than the NGC due to the survey's shallower mean depth in the SGC region.

The data for the NGC and SGC are provided in separate catalogs, each accompanied by a corresponding random catalog. These randoms are $50$ times denser than the data, with redshift distributions generated by randomly shuffling the observed redshifts \citep{2020MNRAS.498.2354R}. This approach introduces a systematic effect known as the Radial Integral Constraint \citep[RIC;][]{deMattia:2019vdg}, the estimation and correction of which is addressed in section~\ref{sec:ic}.

The data and random catalogs include three types of weights assigned to each data point. The first are the close-pair weights, $w_{\text{cp}}$, which correct fiber collisions. The second set of weights, $w_{\text{noz}}$, addresses spectroscopic completeness by accounting for the expected redshift failure rate. Lastly, the imaging systematic weights, $w_{\text{sys}}$, correct systematic effects on large angular scales and are of key importance for $\fnl$ measurements. The total completeness weight assigned to each data point is given by \citep{2020MNRAS.498.2354R}
\begin{equation}
    w_{\text{c}} = w_{\text{cp}} \, w_{\text{noz}} \, w_{\text{sys}} \, .
    \label{eq:wc}
\end{equation}
We remark that this weighting scheme, as expressed in eq.~\eqref{eq:wc}, is also applied to objects in the random catalogs.

To estimate the covariance matrix and the integral constraint corrections to the different summary statistics three sets of $1000$ mock catalogs for each Galactic Cap are provided \citep{2021MNRAS.503.1149Z}. These mocks were generated using the Effective Zel'dovich approximation mock method \citep[EZmock;][]{2015MNRAS.446.2621C} for a flat $\Lambda$CDM cosmology with the following parameters: $\Omega_m = 0.307115$, $\Omega_{\Lambda} = 0.62885$, $\Omega_b = 0.048206$, $h = 0.6777$, $\sigma_8 = 0.8225$, $n_s = 0.9611$, and $\fnl = 0$. The mocks are designed to replicate the two- and three-point clustering statistics of the DR16Q dataset. Figure~\ref{fig:obs-Bk} shows that the observed bispectra (square data points) and the $1 \sigma$ error EZmock error (shaded area) are consistent.

The three EZmocks are the \textit{realistic}, \textit{complete}, and \textit{shuffled} EZmocks. The realistic EZmock set incorporates all known observational systematic effects in its data and random catalogs. Each data catalog in this set has a corresponding random catalog with a shuffled redshift distribution. We use these $1000$ data-random pairs to compute the power spectrum and bispectrum covariance matrix, $\mathbf{\Sigma}$. The other two sets, the complete and shuffled EZmocks, share identical data catalogs free of observational systematic effects. However, their random catalogs differ in how their redshift distributions are produced. In the EZmock shuffled set, each data catalog is paired with a random catalog whose redshift distribution is generated using the shuffling method. By contrast, the EZmock complete set includes only two random catalogs, one for each Galactic Cap, with their redshift distributions sampled from the same smoothed redshift distribution of the EZmock data catalogs. We utilize the EZmock complete and shuffled sets to assess the RIC effect (see section~\ref{sec:ic}).

\subsection{Power spectrum and Bispectrum estimation}\label{sec:estimators}

\begin{figure}
    \centering
    \includegraphics[width=1\linewidth]{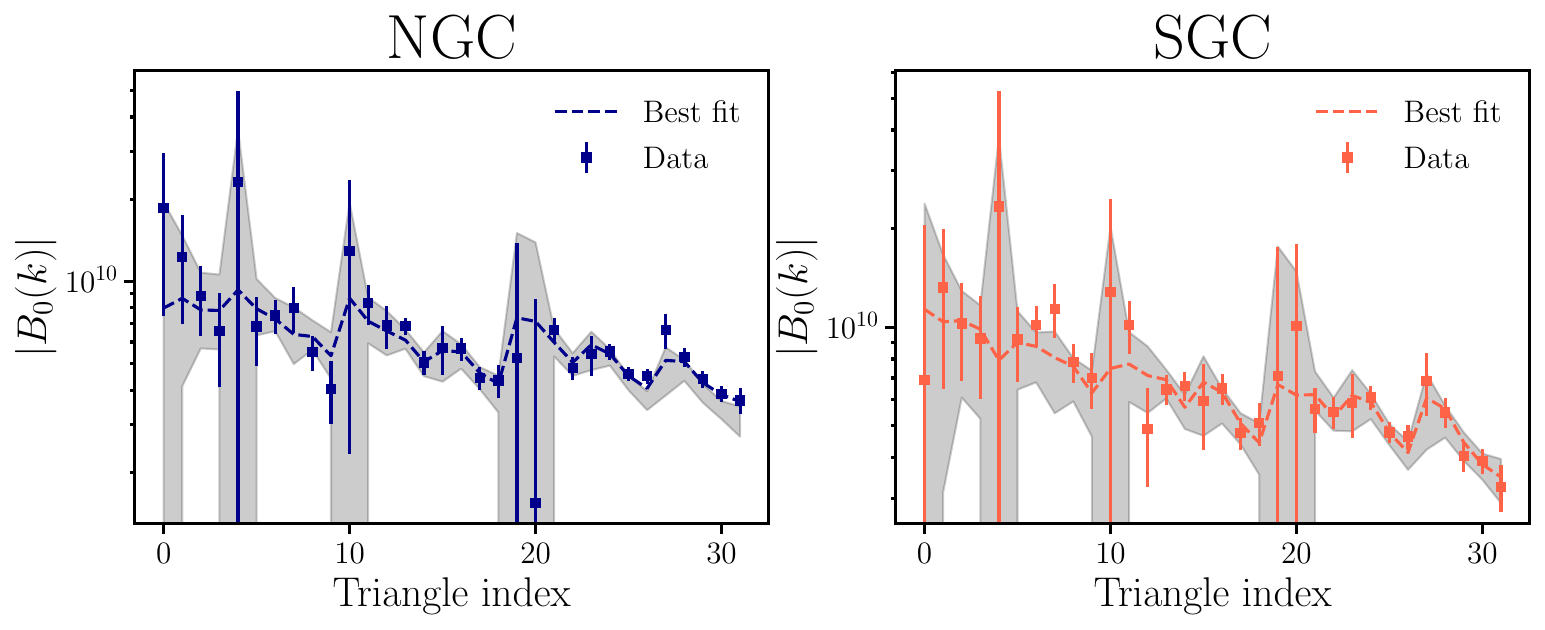}
    \caption{Absolute values of the observed bispectra and best-fit model of the $P+B$ analysis with $p=1.6$ and optimal weights for the power spectrum measurement. We rebinned the data on a $3 \Delta k$ grid. The shaded area corresponds to the $1\sigma$ error of the realistic EZmocks.}
    \label{fig:obs-Bk}
\end{figure}

To write the power spectrum and bispectrum monopole estimators we follow the formalism of ref.~\citep{Scoccimarro:2015bla} and start from the quasar overdensity field that we write in configuration space,
\begin{equation}
    F^w(\mathbf{r}) = w \, \left( w_{\rm c} \, n_{\rm qso} - \alpha_{\rm s} \, w_{\rm c} \, n_{\rm s} \right) \, ,
    \label{eq:F_Cs}
\end{equation}
 and Fourier space,
\begin{equation}
    F_0^w(\mathbf{k}) = \left( \sum_{j}^{N_{\rm qso}} - \, \alpha_{\rm s} \sum_{j}^{N_{\rm s}} \right) \, w_{\text{c}, \,j} \, w_j \, e^{i \mathbf{k} \cdot \mathbf{r}_j} \, ,
    \label{eq:F_Fs}
\end{equation}
where $n_{\rm qso}$ and $n_{\rm s}$ respectively are the quasar and random catalog number density, and $w_{\rm c}$ are the corresponding completeness weights, see eq.~\eqref{eq:wc}. The additional weights $w$ that appear in eqs.~\eqref{eq:F_Cs} and \eqref{eq:F_Fs} can be the standard FKP weights \citep{FKP},
\begin{equation}
    w_{\text{FKP}}(z) = \frac{1}{1 + \bar{n}(z) \, P_{\text{fid}}} \, ,
    \label{eq:wfkp}
\end{equation}
or the optimal weights to extract the local PNG information in LSS from \citep{Castorina2019},
\begin{align}
    \tilde{w}_{\text{tot}}(z) & = w_{\text{FKP}}(z) \, \tilde{w}(z) = w_{\text{FKP}}(z) \, \left(  b_1(z) - p \right) \, , \label{eq:wt} \\
    w_{\text{tot}, \, 0}(z) &= w_{\text{FKP}}(z) \, w_0(z) = w_{\text{FKP}}(z) \, D(z) \, \left( b_1(z) + \frac{f(z)}{3} \right) \, . \label{eq:w0}
\end{align}
We note that to compute the weighting scheme of eqs.~\eqref{eq:wt} and \eqref{eq:w0} the authors of ref.~\citep{Castorina2019} assumed the universal relation \cite{2011JCAP...04..006B,Desjacques:2016bnm}
\begin{equation}
    \bfi(z)=2 \, \delta_{\rm c} \, \left(b_1(z)-p\right) \, , \label{eq:bfi}
\end{equation}
with $\delta_{\rm c}=1.686$ the critical density for spherical collapse.
In eq.~\eqref{eq:wfkp}, $\bar{n}(z)$ is the mean density at redshift $z$ and $P_{\text{fid}} = 3 \times 10^4 \, (\text{Mpc}/h)^3$ is the expected power at the scales affected by PNG. In eqs.~\eqref{eq:wt}-\eqref{eq:bfi} appear the fiducial QSO bias for which we assume the model of \citep{biasQSO},
\begin{equation}
    b_1(z) = 0.278 \, \left( (1 + z)^2 - 6.565 \right) + 2.393 \, ,
    \label{eq:bz}
\end{equation}
the growth factor, $D(z)$, the growth rate, $f(z)$, as functions of redshift, and the response of the QSOs to $\fnl$, $p$. In this work, we use two possible values of $p$: either $p=1.0$, which comes from the universal mass function in spherical collapse \citep{Slosar:2008hx,Biagetti:2019bnp}, or $p=1.6$, which corresponds to recently merged dark matter halos in simulations \citep{Biagetti:2019bnp,Barreira:2020kvh}. \footnote{We note that in the optimal power spectrum analysis we consistently use in the model the same value of $p$ utilized in the optimal estimator, see section~\ref{sec:pbk-model}.}
The final element in eqs.~\eqref{eq:F_Cs} and \eqref{eq:F_Fs} is
\begin{equation}
    \alpha_{\text{s}} = \frac{\sum^{\text{qso}} w_{\text{c}}}{\sum^{\text{s}} w_{\text{c}}} \, ,
    \label{eq:alpha}
\end{equation}
which properly normalizes the random catalog density.

Given the monopole overdensity in Fourier space (eq.~\ref{eq:F_Fs}) the standard, shot-noise subtracted, FKP power spectrum is \citep{Yamamoto2006,Scoccimarro:2015bla}
\begin{equation}
    \widetilde{P}_0^{\rm FKP}(k) = \frac{1}{I_{22}} \, \int \frac{\de \Omega_k}{4 \pi} \, |F_0^{\rm FKP}(\mathbf{k})|^2 - \, S_0^{P^{\rm FKP}} \, ,
    \label{eq:PwFKP}
\end{equation}
while the optimal power spectrum \citep{Castorina2019} for the $\fnl$ signal is the cross-correlation of the overdensities weighted with $\tilde{w}_{\text{tot}}$ and $w_{\text{tot}, \, 0}$ (eqs.~\ref{eq:wt} and \ref{eq:w0}),
\begin{equation}
    \widetilde{P}_0^{\rm Opt}(k) = \frac{1}{A_0} \, \int \frac{\text{d}\Omega_k}{4 \pi} \, F_0^{\tilde{w}}(\mathbf{k}) \, F_0^{w_0}(-\mathbf{k}) - \, S_0^{P^{\rm Opt}} \, ,
    \label{eq:Pwopt}
\end{equation}
where $I_{22}$ and $A_0$ are the normalization factors, and $S_0^{\rm FKP}$ and $S_0^{\rm Opt}$ are the power spectrum shot noises. The normalization factors of the standard and optimal powers spectra read as
\begin{align}
    I_{22} &= \int \de \mathbf{r} \, w_{\rm FKP}^2(\mathbf{r}) \, \left[ w_{\rm c}(\mathbf{r}) \, n_{\rm qso} (\mathbf{r}) \right]^2 \, , \label{eq:int-I22} \\
    A_0 & = \int \text{d}\mathbf{r} \, w_{\text{tot}, \, 0}(\mathbf{r}) \, \tilde{w}_{\rm tot}(\mathbf{r}) \, \left[ w_{\text{c}}(\mathbf{r}) \, n_{\text{qso}}(\mathbf{r}) \right]^2 \, ,
    \label{eq:int-A}
\end{align}
while the shot noise contributions are defined as follows:
\begin{align}
    S_0^{P^{\rm FKP}} &= \frac{1}{I_{22}} \, \int \text{d} \mathbf{r} \, w_{\text{c}}^2(\mathbf{r}) \, n_{\text{qso}}(\mathbf{r}) \, \left( 1 + \alpha_{\text{s}} \right) \, w_{\rm FKP}^2(\mathbf{r}) \, , \label{eq:int-S0PwFKP} \\
    S_0^{P^{\rm Opt}} &= \frac{1}{A_0} \, \int \text{d} \mathbf{r} \, w_{\text{c}}^2(\mathbf{r}) \, n_{\text{qso}}(\mathbf{r}) \, \left( 1 + \alpha_{\text{s}} \right) \, w_{\text{tot}, \, 0}(\mathbf{r}) \, \tilde{w}_{\text{tot}}(\mathbf{r}) \, .
    \label{eq:int-S0PwOpt}
\end{align}
We compute the shot noise contributions from eqs.~\eqref{eq:int-S0PwFKP} and \eqref{eq:int-S0PwOpt} as discrete sums 
\begin{align}
    S_0^{P^{\rm FKP}} &= \frac{1}{I_{22}} \left[ \sum^{N_{\text{qso}}}_j w^2_{\text{c}, \, j} \, w_{\text{FKP}, \, j}^2 + \alpha^2_{\text{s}} \sum^{N_{\text{s}}}_j w^2_{\text{c}, \, j} \, w_{\text{FKP}, \, j} \right] \, , \label{eq:S0PwFKP} \\
    S_0^{P^{\rm Opt}} &= \frac{1}{A_0} \left[ \sum^{N_{\text{qso}}}_j w^2_{\text{c}, \, j} \, w_{\text{tot}, 0, \, j} \, \tilde{w}_{\text{tot}, \, j} + \alpha^2_{\text{s}} \sum^{N_{\text{s}}}_j w^2_{\text{c}, \, j} \, w_{\text{tot}, 0, \, j} \, \tilde{w}_{\text{tot}, \, j}\right] \, .
    \label{eq:S0PwOpt}
\end{align}
As for the normalization factors in eqs.~\eqref{eq:int-I22} and \eqref{eq:int-A}, we approximate them with the limit of the monopole of the window function at small separations, $Q_0(0)$, see section~\ref{sec:meas-window}. For more details, we refer to \citep{Cagliari:2023mkq}.
The estimated power spectra are therefore
\begin{align}
    \hat{P}_0^{\rm FKP}(k) &= \frac{I_{22}}{Q_0^{\rm FKP}(0)} \left( \widetilde{P}_0^{\rm FKP}(k) - S_0^{P^{\rm FKP}} \right) \, ,    \label{eq:P0FKP-norm} \\
    \hat{P}_0^{\rm Opt}(k) &= \frac{A_0}{Q_0^{\rm Opt}(0)} \left( \widetilde{P}_0^{\rm Opt}(k) - S_0^{P^{\rm Opt}} \right) \, ,    \label{eq:P0opt-norm}
\end{align}
where $I_{22}$ and $A_0$ are the discrete versions of eqs.~\eqref{eq:int-I22} and \eqref{eq:int-A},
\begin{align}
     I_{22} &= \alpha_{\rm s} \sum_j^{N_{\rm s}} n_{{\rm s}, \, j} \, w_{\text{c}, \, j}^2 \, w_{\text{FKP}, \, j}^2 \, , \label{eq:I22} \\
    A_0 &= \alpha_{\rm s} \sum_j^{N_{\rm s}} n_{{\rm s}, \, j} \, w_{\text{c}, \, j}^2 \, \tilde{w}_{\text{tot}, \, j} \, w_{\text{tot}, 0, \, j}\, . \label{eq:A0}
\end{align}
Operatively, we measure the standard and optimal power spectra with \texttt{nbodykit} \citep{nbodykit} on a grid of $512^3$ in the smallest non-cubic box containing the survey and assuming Planck \citep{Planck18} as fiducial cosmology.\footnote{The NGC and SGC boxes have the following edge sizes $(3421, 7162, 3012) \, {\rm Mpc}/h$ and $(2658, 5782, 2965) \, {\rm Mpc}/h$ respectively, which were automatically chosen by \texttt{nbodykit}.} The linear grid over which we estimate the power spectra goes from $k_{\text{min}} = 3.75 \times 10^{-3} \, (\text{Mpc}/h)^{-1}$ to $k_{\text{max}} = 2.23 \times 10^{-1} \, (\text{Mpc}/h)^{-1}$ for NGC, and $k_{\text{max}} = 2.78 \times 10^{-1} \, (\text{Mpc}/h)^{-1}$ for SGC.\footnote{We note that \texttt{nbodykit} discretizes eqs.~\eqref{eq:int-I22} and \eqref{eq:int-A} assuming that densities are already weighted, as follows: $$I_{22}^{\rm nb} = \alpha_{\rm s} \sum_j^{N_{\rm s}} n_{\rm s} \, w_{\text{c}, \, j} \, w_{\text{FKP}, \, j}^2 \, , \quad A_0^{\rm nb} = \alpha_{\rm s} \sum_j^{N_{\rm s}} n_{\rm s} \, w_{\text{c}, \, j} \, \tilde{w}_{\text{tot}, \, j} \, w_{\text{tot}, 0, \, j}\, .$$ Therefore, when we re-normalize the power spectrum estimators (eqs.~\ref{eq:P0FKP-norm} and \ref{eq:P0opt-norm}) we multiply them by the factors $I_{22}^{\rm nb}/Q_0^{\rm FKP}(0)$ or $A_0^{\rm nb}/Q_0^{\rm Opt}(0)$.}

On the other hand, for the bispectrum estimation, we only use the FKP weighting scheme. In this case, the bispectrum estimator reads as follows \citep{Scoccimarro:2015bla}:
\begin{equation}
    \widetilde{B}_0 (k_1, k_2, k_3) = \prod_{i=1}^{3} \int_{k_i} \de^3 q_i \, \frac{\delta_D(\mathbf{q}_{123})}{N_{123}^T \, I_{33}} \, F_0^{\rm FKP}(\mathbf{q}_1) \, F_0^{\rm FKP}(\mathbf{q}_2) \, F_0^{\rm FKP}(\mathbf{q}_3) - S_0^B \, ,
    \label{eq:B0-est}
\end{equation}
where $\delta_D$ is the Dirac delta, $N_{123}^T$ is the number of triangles in a bin defined as
\begin{equation}
    N_{123}^T = \prod_{i=1}^{3} \int_{k_i} \de^3 q_i \, \delta_D(\mathbf{q}_{123}) \, ,
    \label{eq:trinagles}
\end{equation}
$I_{33}$ is the bispectrum normalization,
\begin{equation}
    I_{33} = \int \de \mathbf{r} \, w_{\rm FKP}^3(\mathbf{r}) \, \left[ w_{\rm c}(\mathbf{r}) \, n_{\rm qso} (\mathbf{r}) \right]^3 \, ,
    \label{eq:int-I33}
\end{equation}
and $S_0^B$ is the scale-independent shot noise given by
\begin{equation}
    S_0^B = \frac{1}{I_{33}} \, \int \text{d} \mathbf{r} \, w_{\text{c}}^3(\mathbf{r}) \, n_{\rm d}(\mathbf{r}) \, \left( 1 - \alpha^2_{\rm s} \right) \, w_{\rm FKP}^3(\mathbf{r}) \, ,
    \label{eq:int-S0b}
\end{equation}
which in its discrete form becomes
\begin{equation}
    S_0^B = \frac{1}{I_{33}} \left[ \sum^{N_{\rm qso}}_j w^3_{\text{c}, \, j} \, w_{{\rm FKP}, \, j}^3 - \alpha^3_{\rm s} \sum^{N_{\rm s}}_j w_{\text{c}, \, j}^3 \, w_{{\rm FKP}, \, j}^3 \right] \, . \label{eq:S0b}
\end{equation}
As for the scale-dependent shot noise contribution, instead of removing it from the observed bispectrum, we add it to the bispectrum model with a stochastic counterterm whose prior in the analysis is centered on the Poissonian value (see eq.~\ref{eq:B-stoc}).

In the case of the bispectrum, we cannot approximate its normalization factor (eq.~\ref{eq:int-I33}) with the small-scale limit of its window function: therefore, to correctly re-normalize the observed bispectrum we define the normalization correction factor
\begin{equation}
    \beta = \frac{Q_0(0)}{I_{22}} \, .
    \label{eq:beta-corr}
\end{equation}
Then, we compute the normalization of eq.~\eqref{eq:int-I33} using the discrete sum
\begin{equation}
    I_{33} = \alpha_{\rm s} \sum_j^{N_{\rm s}} n_{{\rm s}, \, j}^2 \, w_{\text{c}, \, j}^3 \, w_{\text{FKP}, \, j}^3 \, ,
    \label{eq:I33}
\end{equation}
and we correct it as follows:
\begin{equation}
    I_{33}^{\rm C} = \beta^2 \, I_{33} \, .
    \label{eq:I33-corr}
\end{equation}
We use eq.~\eqref{eq:I33-corr} to normalize the estimators of eqs.~\eqref{eq:B0-est} and \eqref{eq:S0b}.

To compute the bispectrum we used \texttt{Rustico} \cite{GilMarinEtAl2020} on a $512^3$ grid, employing the Piecewise Cubic Spline (PCS) mass assignment scheme and grid interlacing \cite{SefusattiEtAl2015}. For NGC we set $L_\mathrm{box} = 6700 \,\text{Mpc}/h$, while for SGC we use $L_\mathrm{box} = 6400 \,\text{Mpc}/h$. 
We bin the bispectrum in $\Delta k = 5 \times 10^{-3} \, (\text{Mpc}/h)^{-1}$, starting from the bin centered at $k_{\rm min} =  5 \times 10^{-3} \, (\text{Mpc}/h)^{-1}$ to the last bin centered at $k_{\rm max} = 8.0 \times 10^{-2} \, (\text{Mpc}/h)^{-1}$.
We measured the bispectra of the EZmocks using the same configuration but with a $256^3$ grid.
We considered also bins for which the centers do not form a closed triangle, but which do contain closed triangles within the bin.
To reduce the number of bins/triangles analyzed, we also rebin the measured bispectrum monopole into triangles with bin size $N\,\Delta k$ with $N>1$. Let $\{k_1^m, k_2^m, k_3^m\}$, with $m=1, 2, \cdots, N_\mathrm{tri}$ and $N_\mathrm{tri}$ the total number of triangles, denote the center of the triangles in the original measurements, then the rebinned bispectrum $B^{N \Delta k}_0$, defined on the new center of triangles $\{\tilde{k}_1^n, \tilde{k}_2^n, \tilde{k}_3^n\}$ ($n=1, 2, \cdots, \tilde{N}_\mathrm{tri}$) is obtained as 
\begin{equation}
    B^{N \Delta k}_0(\tilde{k}^n_1, \tilde{k}^n_2, \tilde{k}^n_3) = \frac{\sum_{k^m_i \in \tilde{k}^n_i} \, B_0(k^m_1, k^m_2, k^m_3) \, N^T_{123}(k^m_1, k^m_2, k^m_3)}{\sum_{k^m_i \in \tilde{k}^n_i} \,N^T_{123}(k^m_1, k^m_2, k^m_3)} \, ,
\end{equation}
where the summation $\sum_{k^m_i \in \tilde{k}^n_i}$ runs over all triangle centers $\{k^m_1, k^m_2, k^m_3\}$ that satisfy $ \tilde{k}^n_i - N \Delta k \,/2 \leq k^m_i \leq \tilde{k}^n_i + N \Delta k \,/2$ for $i=1,2,3$. 
In figure~\ref{fig:obs-Bk} we show the observed bispectra rebinned on a $3 \Delta k$ grid for NGC (left panel) and SGC (right panel). We also over-plot the best-fit model of the $P+B$ analysis with $p=1.6$ and the mean and 1-$\sigma$ scatter of the realistic EZmocks.

\subsection{Measurements of the window functions}\label{sec:meas-window}

The final data product that we require to compare the observed power spectrum and bispectrum with the model is the window function, $W(\mathbf{s})$, and its multipoles, $Q_{\ell}(s)$. The window function encodes the information of the survey footprint on the sky surface and in redshift.

We perform the measurement of the window function multipoles, which are defined as follows:
\begin{align}
    Q_{\ell}(s) & \equiv (2 \ell + 1) \, \int \text{d} \Omega_s \int \text{d}^3 \mathbf{s}_1 \, W(\mathbf{s}_1) \, W(\mathbf{s} + \mathbf{s}_1) \, \mathcal{L}_{\ell}(\hat{\mathbf{s}}_1 \cdot \hat{\mathbf{s}}) 
     \equiv \int \text{d} s_1 \, s_1^2 \, Q_{\ell}(s; s_1) \, ,
    \label{eq:Ql-def}
\end{align}
with $\mathcal{L}_\ell$ the Legendre polynomials, 
following the same recipe of ref.~\citep{Cagliari:2023mkq} that is based on the pair-counting approach introduced by ref.~\citep{2017MNRAS.464.3121W}. First, using the random catalog, we measure the weighted pair counts, $RR^{w}(s, \mu)$, and their multipoles,
\begin{equation}
    RR^{\text{w}}_{\ell}(s) = (2 \ell + 1) \int \text{d}\mu \, RR^{\text{w}}(s, \mu) \, \mathcal{L}_{\ell}(\mu) \, .
    \label{eq:RR-multipoles}
\end{equation}
Then, we re-normalize them to the data catalog density and the shell width,
\begin{equation}
    Q_{\ell}(s) = \frac{RR^{\text{w}}_{\ell}(s)}{4 \pi \, s^3 \, \text{d}\ln{s}} \frac{\left( \sum^{\text{qso}} w_{\text{c}} \right)^2 - \sum^{\text{qso}} w^2_{\text{c}}}{\left( \sum^{\text{s}} w_{\text{c}} \right)^2 - \sum^{\text{s}} w^2_{\text{c}}} \, ,
    \label{eq:window-l}
\end{equation}
with $\text{d}\ln{s} = \frac{s_{n+1} - s_n}{s}$, where $s$ is the center of the $n$-th separation bin. We remark that for the pair counting the random catalog is weighted according to the weighting scheme used in the power spectrum measurement. For more detail on the window multipole measurements, we refer the reader to ref.~\citep{Cagliari:2023mkq}.

\section{Analysis Methods}\label{sec:analysis}

\subsection{Power spectrum and Bispectrum model}\label{sec:pbk-model}

In Fourier space the quasar density contrast expanded up to the second order in perturbation theory is related to the linear matter overdensity field $\delta_{\rm m}$ through \cite{PhysRevD.81.063530, 2018PhR...733....1D, 2018JCAP...12..035D, 2015JCAP...12..043A, 2011JCAP...04..006B, 1998MNRAS.300..747V,Scoccimarro:1999ed}
\begin{equation} \label{eq:deltaqso}
    \delta_{\rm qso}\left({\kv};z\right) = \Zker_1\left(\kv;z\right)\,\delta_{\rm m}(\kv;z) + \int\de^3\qv_1\de^3\qv_2 \, \delta_{\rm D}\left(\kv-\qv_{12}\right) \, \Zker_2\left(\qv_1,\qv_2;z\right) \, \delta_{\rm m}(\qv_1;z) \, \delta_{\rm m}\left(\qv_2;z\right) \, .
\end{equation}
The first-order and second-order redshift-space kernels $\Zker_1$ and $\Zker_2$ depend on the growth rate $f(z)$, the linear and quadratic local biases $b_1$ and $b_2$, the second-order tidal bias $b_{s^2}$ and, in presence of primordial non-Gaussianity, also on the non-Gaussian bias parameters $\bfi$ and $\bfd$:

\begin{equation} \label{eq:Z1kernel}
    \Zker_1\left(\kv_1;z\right)=\left[ b_1(z)+f(z)\,\mu_1^2+\bfi(z)\,\fnl\,\Mk^{-1}(k_1;z) \right] \, ,
\end{equation}
\begin{equation} \label{eq:Z2kernel}
\begin{split}
    \Zker_2\left(\kv_1,\kv_2; z\right)=&\frac{b_2(z)}{2}+b_{s^2}(z)\,s^2\left(\kv_1,\kv_2\right)+b_1\,F_2\left(\kv_1,\kv_2\right)+f(z)\,\mu_3^2\,G_2\left(\kv_1,\kv_2\right)+ \\
    &+\frac{1}{2}\,f(z)\,\mu_3\,k_3\left[\frac{\mu_1}{k_1}\,\Zker_1\left(\kv_2;z\right)+\frac{\mu_2}{k_2}\,\Zker_1\left(\kv_1;z\right)\right]+ \\
    &+\frac{1}{2}\,\bfi(z)\,\fnl\left[\frac{k_1}{k_2}\,\Mk^{-1}\left(k_1;z\right)+\frac{k_2}{k_1}\,\Mk^{-1}\left(k_2;z\right)\right]\mu_{12}+ \\
    &+\frac{1}{2}\,\bfd(z)\,\fnl\left[\Mk^{-1}\left(k_1;z\right)+\Mk^{-1}\left(k_2;z\right)\right] \, , 
\end{split}    
\end{equation}
where $\mu_i=\hat{\kv_i}\cdot\hat{\bm n}$ and $\mu_{12}=\hat{\bm k}_{\bm 1} \cdot \hat{\bm k}_{\bm 2}$ determine the orientation with respect to the line of sight $\hat{\bm n}$ (respectively related to the angle $\theta_i$ between a wavevector $\hat{\kv_i}$ and $\hat{\bm n}$, and the azimuthal angle $\xi$ that describes a rotation of $\kv_1$ around $\kv_2$).  In \cref{eq:Z2kernel} $s^2$ is the traceless part of the shear field, $F_2$ and $G_2$ are the second-order matter density and velocity kernels:
\begin{align}
    &s^2({\bm k_1},{\bm k_2})=\left(\hat{\bm k}_{\bm 1} \cdot \hat{\bm k}_{\bm 2}\right)^2-\frac{1}{3} \, ,\\
    &F_2\left(\kv_1,\kv_2\right)=\frac{5}{7}+\frac{1}{2}\left(\frac{k_1}{k_2}+\frac{k_2}{k_1}\right)\,\mu_{12}+\frac{2}{7}\,\mu_{12}^2 \, , \\
    &G_2\left(\kv_1,\kv_2\right)=\frac{3}{7}+\frac{1}{2}\left(\frac{k_1}{k_2}+\frac{k_2}{k_1}\right)\,\mu_{12}+\frac{4}{7}\,\mu_{12}^2 \, .
\end{align}

Focusing on non-Gaussian contributions, it is possible to relate the bias parameters to the local biases via the relation presented in eq.~\eqref{eq:bfi} for the first-order non-Gaussian bias and
\begin{equation} 
    \bfd(z)=\bfi(z) + 2\left\{\delta_{\rm c}\left[b_2(z)-\frac{8}{21}\left(b_1(z)-1\right)\right]-b_1(z)+1\right\} \,,\label{eq:bfidelta}
\end{equation}
for the second-order one \cite{2011JCAP...04..006B,Desjacques:2016bnm}.
If following the standard universality relation in eqs.~\eqref{eq:bfi} and \eqref{eq:bfidelta}, one would have $p=1$; however, we will also consider the case of $p=1.6$, that was shown to be more suitable to tracers such as quasars that most likely experienced a recent merger history \cite{Slosar:2008hx}. The characteristic scale dependence of PNG imprints on the distribution of dark matter tracers is instead determined by the function $\Mk$, linking the matter density field to the primordial gravitational potential and given by
\begin{equation}
    \Mk(k;z) = \frac{2 \, c^2\,k^2\,T(k)\,D(z)}{3\,\Omega_{\rm m}\,H_0^2} \, ,
\end{equation}
involving the speed of light $c$, the matter density parameter $\Omega_{\rm m}$, and the Hubble parameter $H_0$ (the latter two evaluated at $z=0$). Furthermore, $\Mk$ depends on the matter transfer function $T(k)$ (normalized to 1 at low-$k$), and on the growth factor $D(z)$ (normalized to $(1+z)^{-1}$ in the matter dominated era).

We account for the stochasticity of the quasar density contrast by including  the following contribution \cite{2018JCAP...12..035D, Castorina:2020png, Rizzo:2023Bz}:
\begin{equation}
    \delta_{\rm qso}^{\rm stoc} = \epsilon(\kv;z)+\epsilon_\delta(\kv;z)\,\delta_{\rm m}(\kv;z)+\fnl\,\epsilon_\phi(\kv;z)\,\Mk^{-1}(k;z)\,\delta_{\rm m}(\kv;z) \, , 
\end{equation}
where we introduced the stochastic fields $\epsilon$, $\epsilon_\delta$ and $\epsilon_\phi$.

Given the quasar density contrast we can then define the power spectrum and the bispectrum, which we decided to both model at tree level, since, even for the power spectrum, there appears to be no need to go beyond linear theory \cite{Cagliari:2023mkq}, as the $\fnl$ signal drops at small $k$, that are moreover dominated by redshift errors. The quasar power spectrum is then
\begin{equation}
    P_{\rm qso}(\kv;z) = G^2\left(\kv; \sigma_{\rm FoG}\right)\,\Zker_1^2(\kv;z)\,P_{\rm m}(k;z) + N \, ,
    \label{eq:Pqso}
\end{equation}
where $P_{\rm m}(k;z)$ is the matter power spectrum in real space, $N$ is introduced to take into account for potential residual shot noise after the subtraction performed and the Lorentzian function $G$ is included to describe the damping of the power spectrum due to non-linear redshift-space distortions and redshift errors:
\begin{equation}    
    G\left(\kv; \sigma_{\rm FoG}\right)=\left[1+\frac{\left(k\,\mu\,\sigma_{\rm FoG}\right)^2}{2}\right]^{-1} \, ,
\end{equation}
with the redshift-independent parameter $\sigma_{\rm FoG}$ accounting both for the quasar velocity dispersion and their redshift error ($\sigma_z=300\,{\rm km \,s^{-1}}$) \cite{2020ApJS..250....8L}.

The bispectrum instead reads
\begin{equation}
    \begin{split}
        B_{\rm qso}\left({\bm k_1},{\bm k_2};z\right)=&\bigl\{\Zker_1\left({\bm k_1};z\right)\,\Zker_1\left({\bm k_2};z\right)\,\Zker_1\left({\bm k_3};z\right)B_{\rm m}\left({\bm k_1},{\bm k_2};z\right)+\\
        &+2\,\Zker_1\left({\bm k_1}\right)\,\Zker_1\left({\bm k_2}\right)\,\Zker_2\left({\bm k_1},{\bm k_2}\right)\,P_{\rm m}\left(k_1;z\right)\,P_{\rm m}\left(k_2;z\right)+2\,{\rm perms} \bigr\} \\
        &+B_{\rm stoc}\left({\bm k_1},{\bm k_2};z\right) \, ,
    \end{split}
    \label{eq:Bqso}
\end{equation}
where $B_{\rm m}\left({\bm k_1},{\bm k_2};z\right)$ is the primordial matter bispectrum, non-vanishing only in the presence of PNG, and the stochastic contribution, allowing for deviations from a Poissonian distribution, is given by
\begin{equation}
    B_{\rm stoc}\left({\bm k_1},{\bm k_2};z\right)=\left\{ \frac{1}{\bar{n}}\left[ (1+\alpha_1)(b_1+\bfi\,\fnl\,\Mk(k_1) \right]\Zker_1({\bm k_1})\,P_{\rm m}(k_1)  + {\rm 2 \, perms} \right\}+\frac{1+\alpha_2}{\bar{n}^2} \, .
    \label{eq:B-stoc}
\end{equation}

For the bispectrum, the stochastic term has both a scale-independent and a scale-independent contribution and, since we subtracted the scale-independent shot noise from the measurements, the $\alpha_2$ parameter plays the role of the parameter $N$ in the power spectrum, accounting for any residuals after the subtraction.

As in the power spectrum, we account for Finger of God corrections in the bispectrum and this is done by including a $\mu$-dependent counterterm in the first order kernel \cite{Ivanov:2022Bprec}, that is modified to
\begin{equation}
    \Zker_1\left(\kv;z\right)\to\Zker_1^{\rm FoG}\left(\kv;z\right)=\left[ b_1(z)+f(z)\,\mu^2+\bfi(z)\,\fnl\,\Mk^{-1}(k;z) -c_1\,\mu^2 \left(\frac{k}{k_{\rm NL}}\right)^2 \right] \, ,
\end{equation}
where $k_{\rm NL}=0.3\,h{\rm /Mpc}$ and the value of $c_1$ depends on the velocity dispersion of the sample.

Finally, it should be mentioned that in the modeling of the bispectrum we chose to keep only PNG contributions that are linear in $\fnl$, as the characteristic scale dependence makes higher-order terms negligible.

Starting from the expressions  of the power spectrum and the bispectrum in eqs.~\eqref{eq:Pqso} and \eqref{eq:Bqso}, we can compute their multipoles as
\begin{align}
    & P_{\ell, {\rm qso}}\left(\kv;z\right) = \frac{2\ell+1}{2}\int_{-1}^1 \,\de\mu\,P_{\rm qso}\left(\kv;z\right)\,\mathcal{L}_\ell(\mu) \, , \label{eq:Pell_qso} \\
    & B_{\ell\mathit{m}, {\rm qso}}\left(\kv_1, \kv_2 ;z\right) = \int_{-1}^1 \,\de\cos\theta_1\,\int_0^{2\pi} \, \de\xi \, B_{\rm qso}\left(\kv_1,\kv_2;z\right) \,Y_\ell^\mathit{m}\left(\theta_1, \xi \right) \, , \label{eq:Bell_qso_lm}
\end{align}
with $Y_\ell^\mathit{m}\left(\theta_1, \xi \right)$ the spherical harmonics. Note that for the bispectrum, with minimal loss of information \cite{2017MNRAS.467..928G, 2022MNRAS.tmp.2216B}, we will only consider multipoles of order $\mathit{m}=0$, whose expression is simplified to 
\begin{equation}
    \begin{split}  
        B_{\ell, {\rm qso}}\left(\kv_1,\kv_2;z\right) & = \sqrt{\frac{2\ell+1}{4\pi}}\,B_{\ell 0, {\rm qso}}\left(\kv_1,\kv_2;z\right) = \\
        & = \frac{2\ell+1}{2} \int_{-1}^1 \,\de\cos\theta_1\, \left[ \frac{1}{2\pi}\int_0^{2\pi} \, \de\xi \, B_{\rm qso}\left(\kv_1,\kv_2;z\right) \right] \mathcal{L}_\ell\left(\cos\theta_1\right) \, .
    \end{split} \label{eq:Bell_qso}
\end{equation}

For our analysis, we only focus on the monopoles $P_0$ and $B_0$, which carry most of the information, but to perform the window convolution, as described in the next section, higher even multipoles of the power spectrum need to be evaluated as well.

\subsection{Window function and effective redshift}\label{sec:window-zeff}

We write the power spectrum convolution with the window function as in ref.~\citep{Cagliari:2023mkq},
\begin{equation}
    P_0(k; z_{\text{eff}}) = \sum_{\ell, \, L} i^{\ell} \begin{pmatrix}
\ell & L & 0\\
0 & 0 & 0
\end{pmatrix}^2 \int \frac{\text{d} q}{2 \pi^2} \, q^2 \, P_{\ell, \, \text{qso}}(q; z_{\text{eff}}) \, Q_{\ell, \, L}(k, q) \, ,
   \label{eq:P0-qso}
\end{equation}
where $P_{\ell, \, \text{qso}}(q; z_{\text{eff}})$ is the $\ell$-th multipole of the QSO power spectrum (eq.~\ref{eq:Pqso}) evaluated at the sample effective redshift $z_{\rm eff}$ (see discussion below), $\big(\begin{smallmatrix}
  \ell & L & 0\\
0 & 0 & 0
\end{smallmatrix}\big)$ is a Wigner 3-$j$ symbol, and
\begin{equation}
    Q_{\ell, \, L}(k, q) \equiv \int \text{d} s \, s^2 j_0(ks) \, j_{\ell}(qs) \, Q_L(s) \, ,
    \label{eq:QlLkp}
\end{equation}
are the mixing matrices defined from the window function multipoles (see eq.~\ref{eq:Ql-def}) and $j_{\ell}$ are the spherical Bessel's functions of order $\ell$. We solve the integral in eq.~\eqref{eq:P0-qso} as a matrix multiplication \citep[see][for details]{Cagliari:2023mkq}.

We performed the bispectrum-window convolution using the following approximation, first proposed in ref.~\citep{GilMarinEtAl2014},
\begin{equation}
\label{eq:B0_conv_approx}
    \tilde{B}_0[P_{\rm m}(k)](\kv_1, \kv_2) \simeq B_0[\tilde{P}_{\rm m}(k)](\kv_1, \kv_2),
\end{equation}
where we used the window-convolved linear matter power spectrum as the input for the bispectrum calculation
\begin{equation}
    \tilde{P}_{\rm m}(k) = \int \frac{d^3 q}{2 \pi^2} \,  q^2 \, P_{\rm m}(q) \, Q_{0,0}(k, q).
\end{equation}

Besides that, the bispectrum is measured by averaging over a number of triangle configurations (see eq.~\ref{eq:B0-est}), an operation called binning. To account for these effects, we evaluated the bispectrum at the \textit{effective triangles} 
\begin{equation}
\tilde{k}_i \equiv \prod_{j=1}^3 \int_{k_j} d^3 q_j  \, \frac{\delta_D(\qv_{123})}{N^T_{123}} \, q_j \, , \hspace{1 cm} \mathrm{for~} i = 1,2,3,
\label{eq:tri_eff}
\end{equation}
as provided by $\mathtt{Rustico}$. An alternative approximation is to evaluate the bispectrum at the \textit{sorted effective triangles} \citep{Oddo:2019run} instead 
\begin{equation}
\tilde{k}^\mathrm{sort}_3 \equiv \prod_{j=1}^3 \int_{k_j} d^3 q_j  \, \frac{\delta_D(\qv_{123})}{N^T_{123}} \, \mathrm{max}(q_1, q_2, q_3) \, ,
\end{equation}
and similar definitions for the middle and smallest values $\tilde{k}^\mathrm{sort}_2$ and $\tilde{k}^\mathrm{sort}_1$, respectively, following the hierarchy $\tilde{k}^\mathrm{sort}_1 \leq \tilde{k}^\mathrm{sort}_2 \leq \tilde{k}^\mathrm{sort}_3$. We have checked that both methods show negligible differences compared to the exact bispectrum binning method; therefore, we opt for eq.~\eqref{eq:tri_eff}, as it is numerically the easiest option. 

\begin{figure*}
\centering
    \includegraphics[width=\columnwidth]{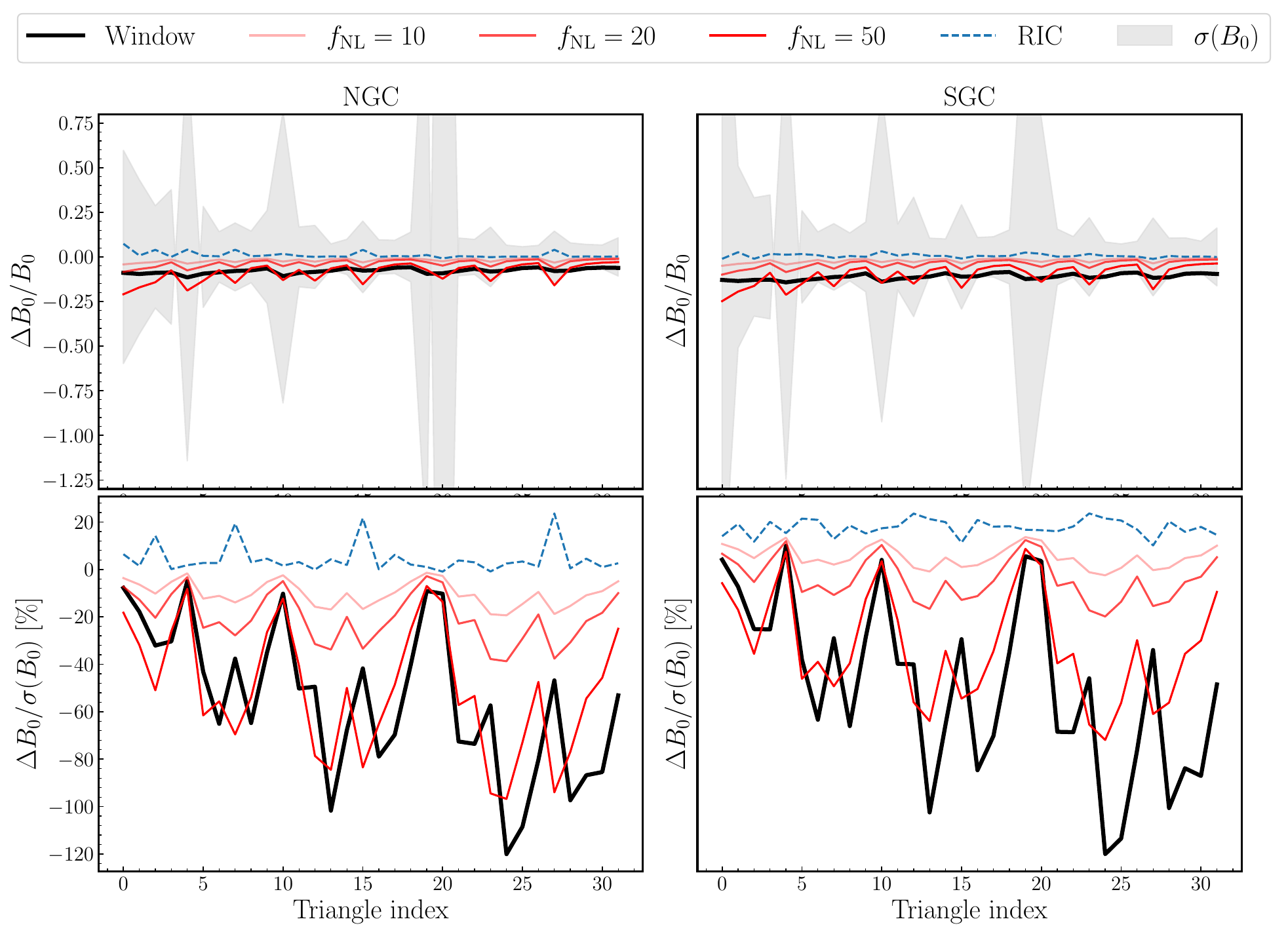}
   \caption{Evaluation of the bispectrum-window convolution approximation. This figure shows the fractional difference between several cases and a fiducial bispectrum monopole model, evaluated with the best-fit bias parameters and $f_\mathrm{NL}=0$, where the bispectrum-window convolution is approximated using eq.~\eqref{eq:B0_conv_approx}. Comparison with the fiducial model (black line) shows that the approximate window treatment introduces relative errors up to $\sim 10\%$, with $>1\sigma$ errors in some triangles. Excluding these triangles has negligible impact on $f_\mathrm{NL}$ constraints. For additional context, we also show the effects of non-zero $f_\mathrm{NL}$ and the inclusion of the RIC.}
     \label{fig:window_check}
\end{figure*}

The accuracy of the bispectrum-window convolution approximation is evaluated in figure~\ref{fig:window_check}. This figure presents the fractional difference between several bispectrum model cases and a fiducial model evaluated at the best-fit bias parameters and $\fnl=0$, with the bispectrum-window convolution treated approximately using eq.~\eqref{eq:B0_conv_approx}. In particular, the black solid line in figure~\ref{fig:window_check} represents the relative difference between the window-convolved bispectrum monopole in eq.~\eqref{eq:B0_conv_approx} and a bispectrum monopole evaluated neglecting the window convolution, $B_0[P_{\rm m}(k)]$, which provides an estimate of the effectiveness of the approximation eq.~\eqref{eq:B0_conv_approx}. The comparison with the fiducial model, computed without accounting for the bispectrum-window convolution (black line), shows that this approximate treatment can introduce a relative error of up to $\sim 10\%$. In some triangle configurations, this corresponds to an error exceeding $1\sigma$. However, we verified that excluding these triangles, which result in errors greater than $1\sigma$, has a negligible impact on the constraints for $\fnl$.

Moreover, from figure \ref{fig:obs-Bk}, we see that our approximate window convolution and effective triangles method provide a good fit to the data, remaining consistent with the error bars. A more stringent test using the mean of the mocks is presented in appendix \ref{app:window_mocks}.

We evaluate the power spectrum and bispectrum models at a fixed effective redshift, $z_{\rm eff}$. Given that the power spectrum and bispectrum estimators in eqs.~\eqref{eq:PwFKP}, \eqref{eq:Pwopt}, and \eqref{eq:B0-est}, contain different powers of the selection function, we define an effective redshift for the power spectrum and one for the bispectrum as follows:
\begin{equation}
    z_{\text{eff}}^P = \frac{\int \de z \, n(z)^2 \,  [\chi(z)^2/H(z)] \, w(z)^2 \, z}{\int \de z \, n(z)^2 \, [\chi(z)^2/H(z)]\,  w(z)^2} \, , \label{eq:zeffP-generic}
\end{equation}
\begin{equation}
    z_{\text{eff}}^B = \frac{\int \de z \, n(z)^3 \,  [\chi(z)^2/H(z)] \, w(z)^3 \, z}{\int \de z \, n(z)^3 \, [\chi(z)^2/H(z)]\,  w(z)^3} \, , \label{eq:zeffB-generic}
\end{equation}
where $n(z)$ is the redshift distribution, $w(z)$ a generic set of weights, $\chi(z)$ the comoving distance, and $H(z)$ the Hubble parameter. The two integrals above are discretized and reduced to
\begin{equation}
    z_{\text{eff}}^P = \frac{\sum^{\text{qso}} z \, n(z) \, w_{\text{c}}^2 \, w_{\text{FKP}}(z)^2 \, \tilde{w}(z) \, w_0(z)}{\sum^{\text{qso}} n(z) \, w_{\text{c}}^2 \, w_{\text{FKP}}(z)^2 \, \tilde{w}(z) \, w_0(z)}\, , \label{eq:zeffP}
\end{equation}
\begin{equation}
    z_{\text{eff}}^B = \frac{\sum^{\text{qso}} z \, n(z)^2 \, w_{\text{c}}^3 \, w_{\text{FKP}}(z)^3}{\sum^{\text{qso}} n(z)^2 \, w_{\text{c}}^3 \, w_{\text{FKP}}(z)^3}\, . \label{eq:zeffB}
\end{equation}
To compute the effective redshift related to the FKP weighted power spectrum we simply remove the optimal weights, $\tilde{w}(z)$ and $w_0(z)$, in eq.~\eqref{eq:zeffP}. For a more detailed discussion about the power spectrum effective redshift approximation accuracy, we refer the reader to figure~4 in ref.~\citep{Cagliari:2023mkq}.

Table~\ref{tab:zeff} summarizes the effective redshift for the power spectrum (top two rows) in the three different weighting configurations, FKP, and optimal weights with $p=1.0$ or $p=1.6$, and the bispectrum (bottom two rows) for the FKP weighting scheme. The effect of the optimal weight is to increase the effective redshift of the sample with respect to the FKP weighting. On the other hand, the bispectrum effective redshift is lower than the corresponding power spectrum one. Not accounting for this difference, i.e. using $z_{\text{eff}}^P$ also for the bispectrum, leads to a $6-9\%$ increase in the $\fnl$ signal and thus an artificial reduction of the uncertainty on $f_{\rm NL}$. 
\begin{table}
    \centering
    \begin{tabular}{ccccc} \toprule 
         & &  FKP & $p=1.0$ & $p=1.6$ \\ \midrule
         \multirow{2}{2em}{$z_{\rm eff}^P$} & NGC & 1.49 & 1.65 & 1.76 \\ \cmidrule(l){2-5}
         & SGC & 1.50 & 1.66 & 1.76 \\ \midrule \midrule
         \multirow{2}{2em}{$z_{\rm eff}^B$} & NGC & 1.47 & -- & -- \\ \cmidrule(l){2-5}
         & SGC & 1.48 & -- & -- \\ \bottomrule
    \end{tabular}
    \caption{The effective redshift for the power spectrum and bispectrum in the two sky regions. For the power spectrum, we also present the optimal-weight effective redshift for two values of the QSO response to $\fnl$, $p$.}
    \label{tab:zeff}
\end{table}

\subsection{The Integral Constraint}\label{sec:ic}

We correct both the power spectrum and the bispectrum with the integral constraint effects, which are due to the use of the data themselves to estimate the survey selection function. We correct two integral constraint effects: the Global Integral Constraint \citep[GIC;][]{1991MNRAS.253..307P,2017MNRAS.464.3121W}, which arises when we assume the survey mean density as the true cosmological mean density, and the RIC \citep{deMattia:2019vdg}, which is related to the radial selection function. In the case of eBOSS DR16 the RIC is caused by the shuffling method used to produce the radial distribution of the random catalog. 

At the power spectrum level, we correct the IC effects as described in ref.~\citep{Cagliari:2023mkq}:
\begin{equation}
    P_0^{\text{IC}}(k) = P_0(k) - P_0(0) \, |W(k)|^2 - P_0(k) \, W_{\text{RIC}}(k) \, ,
    \label{eq:P0-IC}
\end{equation}
where $P_0(k)$ is the monopole of the model power spectrum, $|W(k)|^2$ is the Henkel transform of the monopole of the window function, which is normalized to $|W(0)|^2=1$ \citep{2017MNRAS.464.3121W}, and corrects for the GIC. The RIC correction is 
\begin{equation}
    W_{\text{RIC}}(k) = \frac{\bar{P}_{\text{c}}(k) - \bar{P}_{\text{r}}(k)}{\bar{P}_{\text{c}}(k)} \, ,
    \label{eq:Wric_P}
\end{equation}
where $\bar{P}_{\text{c}}(k)$ and $\bar{P}_{\text{r}}(k)$ are the mean power spectra of the complete and shuffled EZmocks (see section~\ref{sec:data-mocks}).

We must include the integral constraint effects in the model of the observed galaxy bispectrum as well. Similarly to the power spectrum, this constraint ensures that the galaxy bispectrum vanishes at $k \rightarrow 0$. Following ref.~\cite{deMattia:2019vdg}, this can be enforced at the overdensity field level by requiring $\langle \delta_{\mathrm{IC}}(\xv) \rangle = 0$. The integral constraint can be then applied to the correlators. Although enforcing it at the field level is more general, it requires higher-order $N$-point correlation functions (PCFs) of the window, which is technically challenging, especially for the bispectrum (where 4PCF and 5PCF estimates are involved). Therefore, we impose the constraint directly at the correlator level, by inserting a window-convolved linear power spectrum which has been (global) integral constrained corrected, $\tilde{P}_\mathrm{m}^\mathrm{GIC}(k)$, as the input for the bispectrum

\begin{align}
\label{bk_gic_main}
    \tilde{B}_0^\mathrm{GIC}[P_\mathrm{m} (k)](\kv_1, \kv_2) &= B_0[\tilde{P}_\mathrm{m}^\mathrm{GIC}(k)](\kv_1, \kv_2) \, .
\end{align}

Additionally, as i the power spectrum, we need to correct for the RIC in the bispectrum. Following the same treatment as for the power spectrum, we first measure the mean bispectrum of both the complete and shuffled EZmock, denoted as $\bar{B}_c(\kv_1, \kv_2)$ and $\bar{B}_r(\kv_1, \kv_2)$, respectively. The integral constrained corrected bispectrum, including the RIC, is then computed as follows:
\begin{align}
\label{bk_ic_main}
    \tilde{B}_0^\mathrm{IC}[P_\mathrm{m}(k)](\kv_1, \kv_2) &= B_0[\tilde{P}_\mathrm{m}^\mathrm{GIC}(k)](\kv_1, \kv_2) - B_0[\tilde{P}_\mathrm{m}(k)](\kv_1, \kv_2) \, W_{B, \, \mathrm{RIC}}(\kv_1, \kv_2),
\end{align}
where
\begin{equation}
    W_{B, \, \mathrm{RIC}}(\kv_1, \kv_2) = \frac{\bar{B}_c(\kv_1, \kv_2) - \bar{B}_r(\kv_1, \kv_2)}{\bar{B}_c(\kv_1, \kv_2)} 
\end{equation}
represents the fractional difference between the mean bispectra of the complete and shuffled EZmock.

We found that the impact of global integral constraints is negligible for the eBOSS DR16Q dataset. However, the effect of the radial integral constraint constitutes a non-negligible fraction of the error bar, as shown in figure~\ref{fig:window_check}.

\section{Analysis} \label{sec:preanalysis}

\subsection{Fisher Information} \label{sec:fisher}

\begin{table}
    \centering
    \begin{tabular}{cccc} \toprule
         Parameter & Prior & Comments & Fiducial \\ \midrule
         $\fnl$ & $\mathcal{U}[-500, 500]$ & & 0 \\ \midrule
         $b_1^P$ & Correlated with $b_1^B$ & see text, each GC & \cref{eq:bz} \\ \midrule
         $\sigma_{\rm FoG}$ & $\mathcal{U}[0, 20]$ & each GC & $3.74$\\ \midrule
         $N$ & $\mathcal{U}[-5000, 5000]$ & each GC & $0$ \\ \midrule
         $b_1^B$ & Correlated to $b_1^P$ & see text, each GC & \cref{eq:bz} \\ \midrule
         $b_2$ & $\mathcal{U}[-4, 4]$ & each GC & \cref{eq:b2rel} \\ \midrule
         $b_{s^2}$ & $\mathcal{U}[-4, 4]$ & each GC & \cref{eq:bs2rel}\\ \midrule
         $c_1$ & $\mathcal{U}[-1, 1]$ & each GC & 0 \\ \midrule
         $\alpha_1$ & $\mathcal{U}[-1.5, 1.5]$ & each GC & $0$ \\ \midrule
         $\alpha_2$ & $\mathcal{U}[-1.5, 1.5]$ & each GC & $-1$ \\ \bottomrule
    \end{tabular}
    \caption{Parameters and priors used in the analysis. In the table, $\mathcal{U}$ refers to a uniform distribution in the given range. The two first-order biases, $b_1^P$ and $b_1^B$, are correlated with a multivariate Gaussian distribution as explained in section~\ref{sec:likelihood}. When running the joint analysis, the parameters marked as `each GC' are unique to the two Galactic caps for a total of $19$ free parameters. In the last column of the table, we report the fiducial values of the parameters used for the Fisher forecast; $b_1^P$, $b_1^B$, $b_2$ and $b_{s^2}$ have different fiducial values depending on the analyzed GC, as a consequence of the different effective redshift.}
    \label{tab:priorsbis}
\end{table}

Prior to the parameter estimation from the data, we performed a forecast aiming to have a theoretical prediction of the expected improvement of the joint analysis of the quasar power spectrum and bispectrum with respect to the power spectrum-only case. We performed the forecast using the Fisher matrix formalism and we considered the same free parameters of the final analysis, ${\bm \theta}= \left\{\fnl, b_1^P, \sigma_{\rm FoG}, N, b_1^B, b_2, b_{s^2}, c_1, \alpha_1, \alpha_2\right\}$, see table~\ref{tab:priorsbis}, and we kept two distinct first-order biases for the power spectrum and the bispectrum, reflecting the difference in the effective redshift. For every bin, the Fisher matrix is defined as 
\be
    \mathbf{\mathsf{F}}_{\alpha\beta}=\left[\mathbf{\partial}_\alpha  \mathbf{T}(\bm \theta) \cdot  \mathbf{\partial}_\beta \mathbf{ T}(\bm \theta) \right]_{{\bm \theta}={\bm \theta_{\rm fid}}} \, ,
\ee
where $\mathbf{\partial}_\alpha  \mathbf{T}$ indicates the differentiation of the fiducial theoretical model $\mathbf{ T}({\bm \theta}={\bm \theta}_{\rm fid})$ with respect to the $\alpha^{\rm th}$ parameter and $\cdot$ is a generalized scalar product, which, given two vectors $\mathbf{A}$ and $\mathbf{B}$ (e.g.\ the theoretical data vector, its derivative, or the data $\mathbf{D}$) and a covariance matrix ${\bm \Sigma}$, reads
\begin{equation}
 \mathbf{A} \cdot  \mathbf{B} \equiv \sum_{i,j}\,A_i^\mathsf{T}\, {\bm \Sigma}_{ij}^{-1}\,B_j,
 \label{eq:gen_scalar_prod}
\end{equation}
where the sum is intended to be over all the $i$, $j$ bins in the vectors. The marginal error on the $\alpha^{\rm th}$ parameter can be estimated from the Fisher matrix as $\sigma_{{\rm \theta}_\alpha}=\sqrt{\left(\mathbf{\mathsf{F^{-1}}}\right)_{\alpha\alpha}}$, which, in light of the Cramér-Rao inequality, is the lower limit on the achievable error bars \citep{cramer,Rao}. For the forecast we compared the $P$-only and the $P+B$ analyses, in the same configurations that we will consider for the parameter estimations: each GC at a time and then NGC and SGC jointly, and using the covariance matrix estimated from the realistic EZmocks, in consistency with what is done for the analysis.

To build the theoretical data vectors we used the fiducial values of the parameters reported in table~\ref{tab:priorsbis}, where we utilized the following bias relations for $b_2$ and $b_{s^2}$ \citep{2016JCAP...02..018L,2009JCAP...08..020M,2012PhRvD..86h3540B}:
\begin{align}
    b_2 &= 0.412-2.143 \, b_1^B + 0.929 \left( b_1^B \right)^2 + 0.008 \left( b_1^B \right)^3 \, , \label{eq:b2rel} \\
    b_{s^2}&=-\frac{2}{7}\left(b_1^B-1\right) \, . \label{eq:bs2rel}
\end{align}
We chose as fiducial value for $\sigma_{\rm FoG}$ the best-fit value in ref.~\cite{Cagliari:2023mkq} and we used the same value for SGC and NGC. We imposed $c_1=0$ since we expect these effects to be subdominant in the bispectrum. To model the shot noise, we used as reference the known mean number density of quasars in each GC and, to be consistent with the shot noise subtraction in the power spectrum and in the bispectrum (from which only the scale-independent part is subtracted as pointed out in \cref{sec:estimators}), we set $N=0$ and $\alpha_2=-1$.\footnote{To avoid numerical instabilities and a singular Fisher matrix, we set the derivative with respect to $N$ to a constant value of $10^{-9}$.} Finally, we considered a fiducial Poissonian scale-dependent shot noise in the bispectrum imposing $\alpha_1=0$.

\begin{table}
\centering
\begin{tabular}{ccrrr}
         & $p$ & & $\sigma_{\fnl}^{P+B}$ & $\Delta_\sigma^{\%}$ \\ \midrule
         \multirow{4}{4em}{joint} & \multirow{2}{1em}{$1.0$} & \multirow{1}{3em}{FKP} & $12.8$ & $7.5\%$ \\ \cmidrule(l){3-5}
         & & \multirow{1}{3em}{Optimal} & $11.4$ & $8.2\%$ \\ \cmidrule(l){2-5}
         & \multirow{2}{1em}{$1.6$} & \multirow{1}{3em}{FKP} & $23.8$ & $7.4\%$ \\ \cmidrule(l){3-5}
         & & \multirow{1}{3em}{Optimal} & $17.2$ & $9.1\%$ \\ 
         \midrule \midrule
         \multirow{4}{4em}{NGC} & \multirow{2}{1em}{$1.0$} & \multirow{1}{3em}{FKP} & $16.2$ & $5.6\%$ \\ \cmidrule(l){3-5}
         & & \multirow{1}{3em}{Optimal} & $14.4$ & $5.6\%$ \\ \cmidrule(l){2-5}
         & \multirow{2}{1em}{$1.6$} & \multirow{1}{3em}{FKP} & $30.3$ & $5.5\%$ \\ \cmidrule(l){3-5}
         & & \multirow{1}{3em}{Optimal} & $22.5$ & $6.0\%$ \\
         \midrule \midrule
         \multirow{4}{4em}{SGC} & \multirow{2}{1em}{$1.0$} & \multirow{1}{3em}{FKP} & $21.0$ & $10.7\%$ \\ \cmidrule(l){3-5}
         & & \multirow{1}{3em}{Optimal} & $18.3$ & $12.6\%$ \\ \cmidrule(l){2-5}
         & \multirow{2}{1em}{$1.6$} & \multirow{1}{3em}{FKP} & $38.7$ & $10.7\%$ \\ \cmidrule(l){3-5}
         & & \multirow{1}{3em}{Optimal} & $28.1$ & $14.0\%$ \\
         \bottomrule
    \end{tabular}
    \caption{$68\%$ error bars on $\fnl$ resulting from the Fisher analysis of $P+B$ and percentage improvement with respect to a $P$-only Fisher forecast, for the different configurations tested; $\Delta_\sigma$ is defined as $\Delta_\sigma=\left|\sigma_{\fnl}^{P}-\sigma_{\fnl}^{P+B}\right| \,/ \,\sigma_{\fnl}^{P}$ and in the last column of the table it is expressed as a percentage.}
    \label{tab:fisher}
\end{table}

The forecasts, summarized in table~\ref{tab:fisher}, indicate that the expected improvement of the $P+B$ analysis with respect to the $P$-only analysis is between $5\%$ and $15\%$ depending on the weighting scheme and the choice of the value of $p$ in the $\bfi(b_1)$ and $\bfd(b_1,b_2)$ relations (eqs.~\ref{eq:bfi} and \ref{eq:bfidelta}), and the tightening of the error bars on $\fnl$ for the joint analysis of the two GC is in between the one for SGC (which leads to the greatest gain) and the one for NGC. 

These tests can serve as a guideline for the following analysis, not only to have a reference point for the foreseen enhancement of the constraining power on $\fnl$ but also to provide guidance for the choice of the priors on the parameters. The size of the priors was initially set according to the results of the Fisher forecast on the marginal uncertainties on each free parameter; however, after testing that this had no effects on the posterior distribution of the parameters of our analysis but was only speeding up the analysis, we decided to put tighter constraints (the ones reported in table~\ref{tab:priorsbis}) on the second order bias, $b_2$, and on the counterterm $c_1$.

\subsection{Bispectrum compression and Shape correlation} \label{sec:compression-cosine}

\begin{figure}
    \centering
    \includegraphics[width=.8\columnwidth]{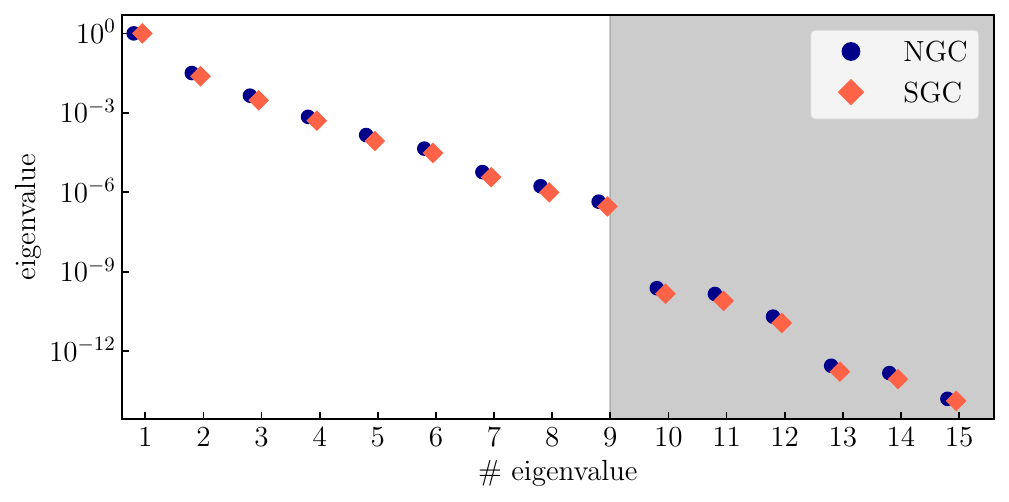}
    \caption{The (normalized) eigenvalues of the matrix $X^T X$, where is $X_{ia}$ is defined in eq.~\eqref{eq:X_matrix}. For the default analyses, we select the top 9 modes corresponding to the largest eigenvalues.}
    \label{fig:eig_vals_SVD}
\end{figure}
In this work, following the method outlined in ref.~\citep{Philcox:2020zyp}, we also consider an analysis of the compressed bispectrum using the singular value decomposition (SVD) algorithm, which allows for the identification of the modes of the bispectrum data vector that contribute the most to the log-likelihood. To begin, we generate $N_\mathrm{bank}$ realizations of (binned) bispectrum $B_a(\bm{\theta})$, where $a = 1, \cdots, N_\mathrm{tri}$, by drawing parameters $\bm{\theta}^i$ from a prior distribution. The bispectrum is computed using the model described in section~\ref{sec:pbk-model} by sampling the galaxy bias, counterterms, and noise parameters from generously large priors.

We then construct an $N_\mathrm{bank} \times N_\mathrm{bins}$ matrix 
\begin{equation}
    X_{ia} \equiv \sum_b \mathcal{S}_{ab}^{-1} \left( B_b(\bm{\theta}^i) - \bar{B}_b \right) \, ,
\end{equation}
where $\bar{B}_b$ is the average over the $N_\mathrm{bank}$ realizations, while $\mathcal{S}_{ab} \equiv \mathcal{C}_{ab}^{1/2}$ represents the square root of the covariance $\mathcal{C}_{ab}$ across the realizations. The covariance is estimated using an analytical prediction under the Gaussian approximation \citep[for details, see e.g., section~4.2 of][]{Rizzo:2023Bz}. To account partially for the survey geometry in the covariance, we use the power spectrum multipoles measurements as input and re-weight the covariance by a volume factor $V_\mathrm{fact} \equiv \left(2\pi/\Delta k\right)^3 \,V_\mathrm{eff}^{-1}$, where $\Delta k$ is the original bin size of the bispectrum measurement, while $V_\mathrm{eff}$ is the effective volume of the samples (see section~\ref{sec:data-mocks}). Finally, performing SVD on the $X_{ia}$ matrix yields 
\begin{equation}
\label{eq:X_matrix}
    X_{ia} = \sum_{\alpha = 1}^{N_\mathrm{tri}} U_{i\alpha} \, D_\alpha \, V_{\alpha a} \, ,
\end{equation}
where $U_{i\alpha}$ is an $N_\mathrm{bank} \times N_\mathrm{tri}$ matrix with orthonormal columns, $D_\alpha$ is an $N_\mathrm{tri}$ vector and $V_{\alpha a}$ is an $N_\mathrm{tri} \times N_\mathrm{bank}$ matrix with orthonormal rows.

We can then construct the bispectrum measurements in the new basis $B_\alpha^\mathrm{SVD}$ 
\begin{equation}
    B^\mathrm{SVD}_\alpha = \sum_{ab} V_{\alpha a} \, \mathcal{S}_{ab} \, B_b \, ,
\end{equation}
where operationally, $V_{\alpha a}$ can be obtained by computing the eigenvectors of $X^T X$. This allows us to limit the analysis to a reduced set of bispectrum modes, selecting them based on the corresponding eigenvalues of $X^T X$. The (normalized) eigenvalues of both galaxy caps are shown in figure~\ref{fig:eig_vals_SVD}. For the default analyses, we select the top 9 modes corresponding to the largest eigenvalues, which also correspond to the `elbow' of the eigenvalue curve.

To make a first comparison between this compression method and the re-binning we quantify the shape difference between a theoretical model $\mathbf{T}(\bm{\theta})$ and the data $\mathbf{D}$ with the \textit{cosine} \citep{Babich:2004gb,Smith_2009}, defined as
\begin{equation}
    \cos \left( \mathbf{T}(\bm \theta), \mathbf{D} \right) \equiv \frac{ \mathbf{T(\bm \theta)} \cdot \mathbf{D}}{\sqrt{\left( \mathbf{T}(\bm \theta) \cdot \mathbf{T}(\bm \theta) \right) \left( \mathbf{D} \cdot \mathbf{D} \right)}}\, ,
\end{equation}
where $\cdot$ represents the generalized scalar product, introduced in eq.\,\eqref{eq:gen_scalar_prod}, which employs the numerical covariance from the mocks. In particular $\sqrt{\mathbf{D} \cdot \mathbf{D}}$ is the usual signal-to-noise ratio (SNR). 

\begin{figure*}
\centering
    \includegraphics[width=\columnwidth]{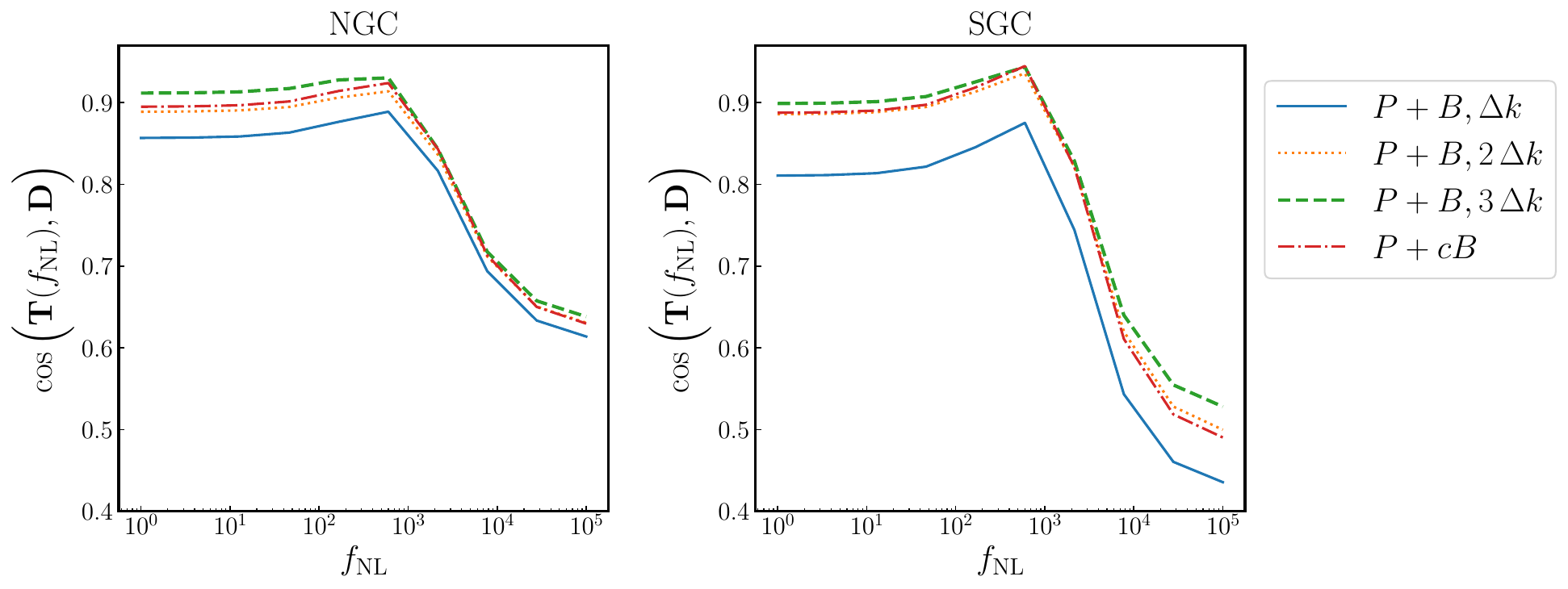}
   \caption{Cosine plot for different bispectrum models and datasets. The highest correlation is observed for data with $3\Delta k$, followed by the compressed bispectrum, in agreement with the $\fnl$ constraint in table~\ref{tab:fnl}.}
     \label{fig:cosine_check}
\end{figure*}

We aim to quantify the shape difference between the $\fnl$ terms of the bispectrum model and the dataset for various bispectrum measurements and data compressions. Specifically, we focus on the non-Gaussian contributions to the bispectrum model for a given $\fnl$, with all other bias parameters fixed to the best-fit values from the MCMC analysis. The results are shown in figure~\ref{fig:cosine_check}. We observed that the cosine is less sensitive to small $\fnl$ variations, while, as expected, the data exhibit weak correlation for models with large $\fnl$. Rebinning the data to a coarser grid, removes some noise from the numerical covariance, leading to a cleaner $\fnl$ signal, as indicated by the increase in the cosine. Additionally, the compression procedure seems to better isolate the $\fnl$ signal, as shown by the high correlation in the cosine plot. Overall, the cosine plot shows the highest correlation for data with $3\Delta k$ bin size, followed by the compressed bispectrum which is broadly consistent with the $\fnl$ constraint presented in table~\ref{tab:fnl}.

\subsection{Parameter estimation}\label{sec:likelihood}

We assume a Gaussian likelihood,
\begin{equation}
    \mathcal{L}(\mathbf{D}|\boldsymbol{\theta}, \mathbf{\Sigma}) \propto \exp \left[ - \frac{1}{2} \, \left(\mathbf{D}-\mathbf{T}({\bm \theta})\right) \cdot \left(\mathbf{D}-\mathbf{T}({\bm \theta})\right) \right] \, ,
    \label{eq:likelihod}
\end{equation}
to sample the posterior distribution of the parameters $\boldsymbol{\theta}$ of model $\mathbf{T}(\boldsymbol{\theta})$, given the data vector $\mathbf{D}$. The free parameters of the analysis with the corresponding prior distributions are reported in table~\ref{tab:priorsbis}. All the other cosmological parameters are fixed to the Planck best-fit values \citep{Planck18}. We expect the first-order biases of the power spectrum and the bispectrum to follow the same redshift evolution and be correlated \citep{DAmico:2022osl}. Therefore, we use as their prior a multivariate Gaussian, $\mathcal{N}(\boldsymbol{\mu}, \mathbf{\Sigma}_b)$, that has mean values
\begin{equation}
    \boldsymbol{\mu} = [2.30, 2.28] \, ,
    \label{eq:mub1}
\end{equation}
which are the expected bias values from eq.~\eqref{eq:bz} at the FKP effective redshifts of the power spectrum and bispectrum. The covariance between the linear bias parameters reads as follows
\begin{equation}
    \mathbf{\Sigma}_b = 
    \begin{pmatrix}
        \sigma_{b_1^P}^2 &  \rho_{12} \, \sigma_{b_1^P} \, \sigma_{b_1^B} \\
        \rho_{12} \, \sigma_{b_1^P} \, \sigma_{b_1^B} & \sigma_{b_1^B}^2
    \end{pmatrix} \, ,
    \label{eq:b1Cov}
\end{equation}
where $\sigma_{b_1^P}^2 = \sigma_{b_1^B}^2 = 2$ are the first-order bias variances and $\rho_{12} = \left( 1 - \varepsilon^2/2 \right)$ is the correlation parameter with $\varepsilon = 0.2$. 

In this work, we analyze two different data vectors in three configurations. The configurations are first of all the separate analysis of the NGC and SCG samples and then their joint analysis. In the joint analysis case, all the biases and nuisance parameters are unique to the two fields of view, while $\fnl$ is shared. The data vectors that we analyze are either the combination of the power spectrum and the bispectrum with a binning of $3 \Delta k$ or the combination of the power spectrum and the compressed bispectrum, see section~\ref{sec:compression-cosine}. We remind the reader that the observed power spectrum can be FKP or optimally weighted.

Due to the high dimensionality of the parameter space ($19$ for the joint analysis), we use a Hamiltonian Monte Carlo (HMC) method~\citep[HMC;][]{1987PhLB..195..216D,2011hmcm.book..113N,2017arXiv170102434B} to run the posterior distribution sampling. HMC is a powerful sampling technique that combines principles from Hamiltonian mechanics and MCMC to navigate the parameter spaces and is especially efficient in high-dimensional problems. Unlike traditional MCMC methods, HMC integrates the target probability distribution dynamically and introduces auxiliary momentum variables, transforming the sampling problem into a Hamiltonian system where the potential energy is defined as the negative logarithm of the joint likelihood. Using the gradients of the probability distribution, HMC follows Hamiltonian trajectories, enabling efficient sampling even in high-dimensional spaces. In this work, we use the No-U-Turn Sampler \citep[NUTS;][]{2011arXiv1111.4246H,ge2018t} version of HMC, which automatically adjusts parameters, such as step size and trajectory length. Setting an acceptance rate of $0.65$ ensures convergence, with the chains achieving $R - 1 < 10^{-3}$, where $R$ is the Gelman-Rubin statistic~\citep{1992StaSc...7..457G}, within a few thousand steps. In each analysis, we run $12$ parallel and independent chains. 

\section{Constraints and Discussion}\label{sec:results}

\begin{figure}
    \centering
    \includegraphics[width=.5\textwidth]{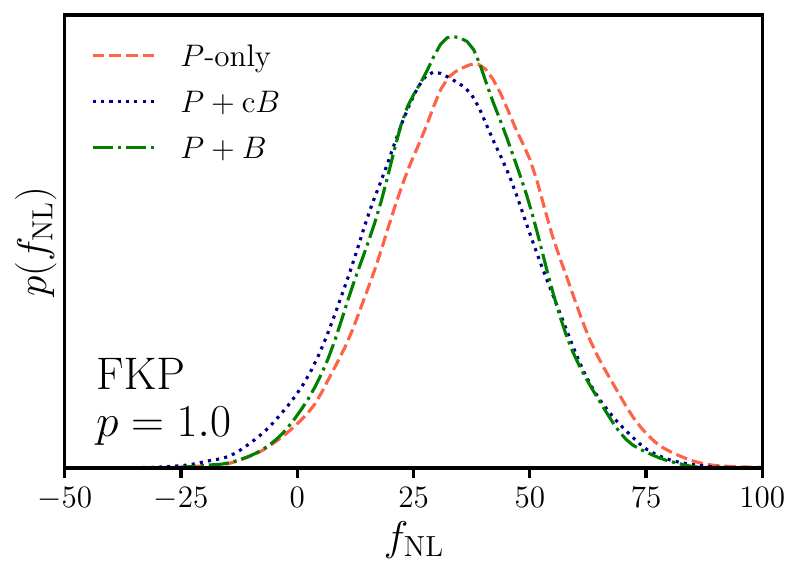}\includegraphics[width=.5\textwidth]{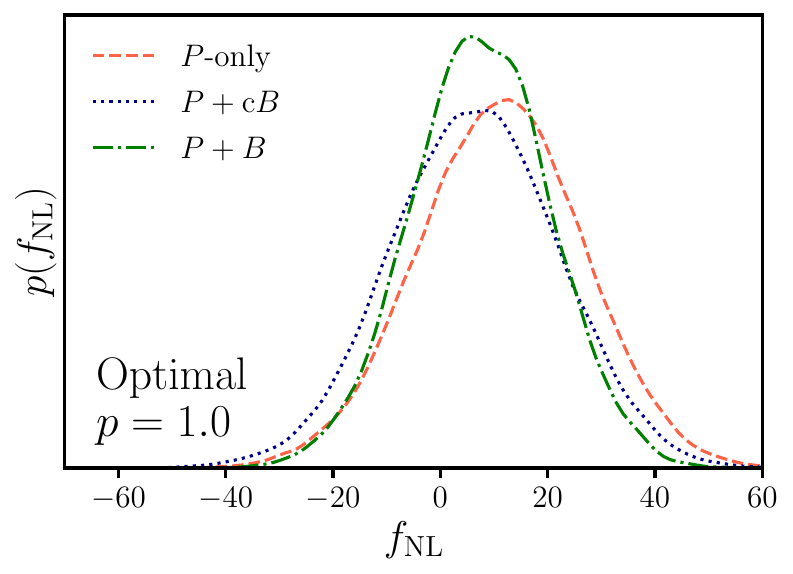}
    \caption{One dimensional posterior distribution for $f_{\text{NL}}$ for the joint analysis of the linear catalog for $p=1.0$. The dashed red curve is the posterior distribution obtained by the power spectrum analysis of ref.~\citep{Cagliari:2023mkq}, the dotted blue curve by the power spectrum plus compressed bispectrum analysis, and the dash-dotted green curve by the power spectrum plus bispectrum analysis. On the left, are the results for the FKP weighted power spectrum, and on the right for the optimally weighted power spectrum.
    }
    \label{fig:fnl-p1p0}
\end{figure}
\begin{figure}
    \centering
    \includegraphics[width=.5\textwidth]{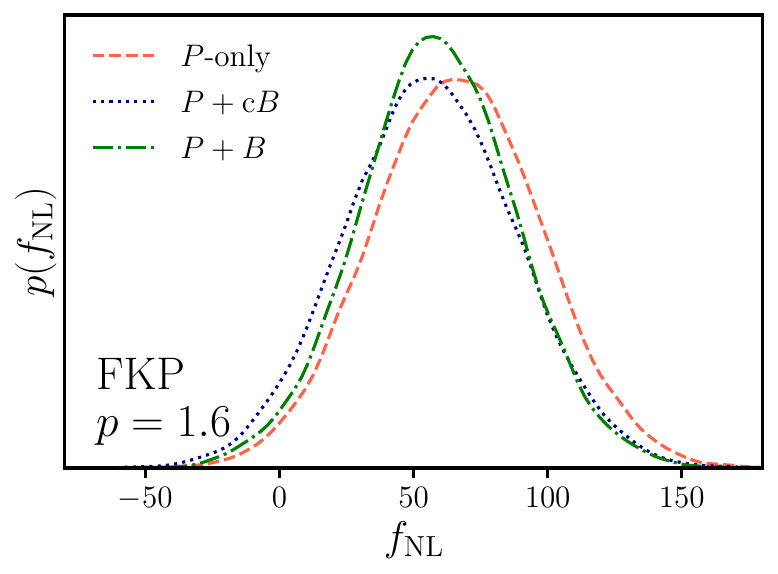}\includegraphics[width=.5\textwidth]{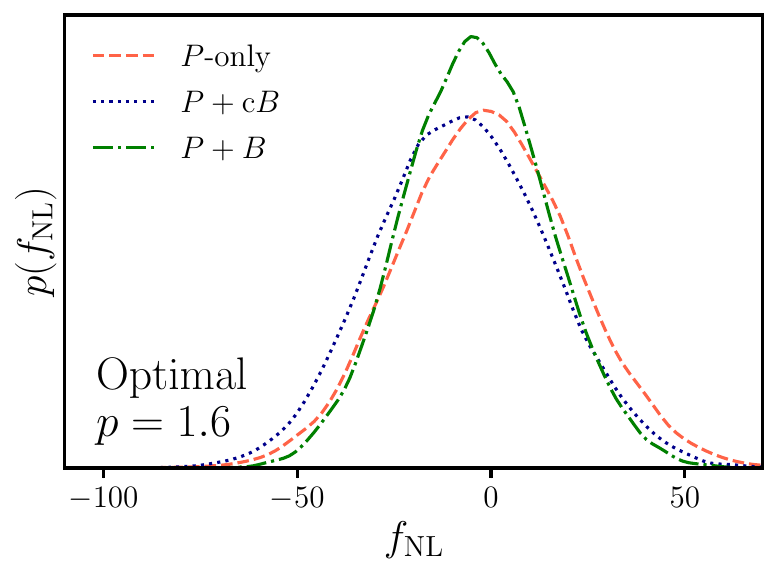}
    \caption{Same as figure~\ref{fig:fnl-p1p0}, but for $p=1.6$.
    }
    \label{fig:fnl-p1p6}
\end{figure}
In this section, we discuss the results of the analyses that combine the power spectrum and bispectrum, which can be either binned in $3 \Delta k$ (hereafter $P+B$) or compressed into $9$ modes (hereafter $P+{\rm c}B$, see section~\ref{sec:compression-cosine}). In figure~\ref{fig:obs-Bk} we present the observed bispectra in the two fields of view overplotted with the $P+B$ with $p=1.6$ and optimal $P$ best fit.

We show our main results in figures~\ref{fig:fnl-p1p0} and \ref{fig:fnl-p1p6}, where we plot the one-dimensional posterior distributions of $\fnl$ for the joint analyses with $p=1.0$ (figure~\ref{fig:fnl-p1p0}) and $p=1.6$ (figure~\ref{fig:fnl-p1p6}). In both figures, the left panel presents the result that combines the FKP weighted power spectrum with the bispectrum, while the right panel presents the optimally weighted power spectrum plus bispectrum constraints. The plots also compare the posterior distribution of the $P+B$ analysis (dash-dotted green line) with the $P+{\rm c}B$ (dotted blue line) and the $P$-only analysis (dashed red line) of ref.~\citep{Cagliari:2023mkq}. In table~\ref{tab:fnl}, we present all the $68\%$ and $95\%$ bounds for the different analysis configurations and combinations of fields of view (NGC, SGC, and joint). 

In all cases, we observe that the $P+B$ analysis improves the constraints with respect to the $P$-only analysis. In particular, for the joint analyses, there is a $6\%$ and $8\%$ improvement when the observed power spectrum is FKP weighted and $p$ is respectively chosen to be $1.0$ or $1.6$. On the other hand, when we use the optimally weighted power spectrum the improvement becomes of the order of $16\%$ for both choices of $p$. These improvements are consistent with the Fisher information analysis. Our final constraints are close to the limit set by the Cramer-Rao inequality (see section~\ref{sec:fisher}), especially when applying optimal weights for the power spectrum ($14\%$ and $6\%$ difference respectively for $p=1.0$ and $p=1.6$ in the joint caps analyses), meaning that our estimator is close to the maximum-likelihood estimator. In addition to tighter constraints, the $P+B$ analyses tend to move the bounds toward $0$ in the FKP power spectrum case, or to negative $\fnl$ in the optimal power spectrum case. Finally, we observe that using the optimal power spectrum in the $P+B$ analysis de-bias the posterior distribution (making them consistent with $\fnl=0$ and CMB measurements). This results in tighter bounds and explains the larger improvement with respect to the $P$-only analysis.

Conversely, the $P+{\rm c}B$ analyses do not produce any improvements with respect to the $P$-only results. In some cases, we even observe a degradation of the constraints. We cannot point to a definitive reason for this behavior but there are at least two possibilities. First, the data compression is not based on $\fnl$-only information: therefore, there is no guarantee that the compressed bispectrum contains the same amount of $\fnl$ information as the full-shape bispectrum. We possibly lose part of the information in the low eigenvalue modes that we discard in the compression. This explanation is backed up by the cosine plot in figure~\ref{fig:cosine_check}, where the $P+{\rm c}B$ curve is always lower than the $P+B$ one. Second, the covariance matrix may be too noisy, especially in those elements that correlate the power spectrum with the compressed bispectrum, and this can degrade the constraining power of the $P+{\rm c}B$ summary statistic.

In table~\ref{tab:fnl} we also present the results for the single field analyses. Similarly to the FKP $P$-only constraints when we combine the power spectrum with the bispectrum (compressed or not) the NGC results are non-consistent with $\fnl=0$ at $95\%$ level; while the SGC constraints always are. Nevertheless, the SGC and NGC constraints are consistent with each other in all cases showing that we can perform the joint analysis. In figures~\ref{fig:corner-FKP} and \ref{fig:corner-Opt} we plot the two-dimensional posterior distributions of the whole parameter space for the $P+B$ and $P+{\rm c}B$ joint analyses with $p=1.0$ for the FKP power spectrum (figure~\ref{fig:corner-FKP}) and the optimally weighted power spectrum (figure~\ref{fig:corner-Opt}). These plots show that the two analyses are consistent for all the parameters and that we have no constraining power over the nuisance parameters $b_2$ and $c_1$ (see section~\ref{sec:fisher}). Finally, we can clearly see from figure~\ref{fig:corner-FKP} that the posterior distribution of the bispectrum first-order bias, $b_1^B$, is centered on smaller values than the power spectrum bias, $b_1^P$. This is consistent with the different choices of effective redshift for the power spectrum and bispectrum models, see table~\ref{tab:zeff}.

Finally, with an eye towards upcoming spectroscopic survey data, we repeated the forecasts for a hypothetical final release of the DESI QSO catalog, assuming a redshift coverage $0.8<z<3.1$ on one-third of the sky, and a number density two times higher than the one of the eBOSS QSO sample. For the power spectrum we take $k_{{\rm min},P} = 2\pi /V^{1/3}  = 1 \times 10^{-3} \, (\text{Mpc}/h)^{-1}$, while for the bispectrum either $k_{{\rm min},B} = 1 \times 10^{-3} \, (\text{Mpc}/h)^{-1}$ or $k_{{\rm min},B} = 3 \times 10^{-3} \, (\text{Mpc}/h)^{-1}$ (conservatively we still fix $k_{\rm max} = 0.1 \,(\text{Mpc}/h)^{-1})$. For these forecasts we assume a Gaussian covariance, nor we can include any systematic effects that could reduce the constraining power of the actual data.

At the level of the power spectrum alone, our hypothetical final DESI catalog could improve up to $87\%$ with respect to our benchmark eBOSS configuration, almost regardless of the assumed value of $p$. This is mostly driven by the larger volume and smaller value of $k_{\rm min}$ compared to eBOSS, with the lower shot noise playing a minor role. In terms of the joint analysis of power spectrum and bispectrum, we forecast that the analysis of the DESI QSO could reach $\sigma (\fnl) \lesssim 1$.  In particular, if the full range of scales is accessible a $P+B$ analysis would improve by $\sim60\%$ over $P$ alone, an improvement which is reduced to  $\sim$32-37\% for the higher value of $k_{{\rm min},B}$.
In practice, the value of $k_{{\rm min}}$ will depend in a number of ways on the sample, as, for example, systematic effects could impact larger scales that might have to be excluded from the analysis. For the bispectrum, the binning of the data also matters. On the one hand, binning is necessary to reduce the number of triangle configurations and make the problem computationally tractable. At the same time though, the binning procedure not only increases the value of $k_{{\rm min},B}$, but it also removes squeezed triangles from the dataset, which contain most of the local PNG signal. We envisage that different binning strategies on different scales, finer/larger bins at lower/higher $k$, might recover most of the signal while keeping the dimensionality of the problem sufficiently small, but this approach would have to be tested in mocks and data.

\begin{table}
    \centering
    \begin{tabular}{ccrrrrr} \toprule
         & $p$ & & C.L. & $P+B$ & $P+\text{compressed} \, B$ & $P$-only\\ \midrule
         \multirow{8}{2em}{joint} & \multirow{4}{1em}{$1.0$} & \multirow{2}{3em}{FKP} & $68\%$ & $18 < f_{\text{NL}} < 50$ & $15 < f_{\text{NL}} < 50$ & $19 < f_{\text{NL}} < 53 $\\ \cmidrule(l){4-7}
         & & & $95\%$ & $3 < f_{\text{NL}} < 66$ & $-2 < f_{\text{NL}} < 67$ & $4 < f_{\text{NL}} < 71$ \\ \cmidrule(l){3-7}
         & & \multirow{2}{3em}{Optimal} & $68\%$ & $-6 < f_{\text{NL}} < 20$ & $-9 < f_{\text{NL}} < 22$ & $-4 < f_{\text{NL}} < 27$ \\ \cmidrule(l){4-7}
         & & & $95\%$ & $-18 < f_{\text{NL}} < 33$ & $-24 < f_{\text{NL}} < 37$ & $-18 < f_{\text{NL}} <42$  \\ \cmidrule(l){2-7}
         & \multirow{4}{1em}{$1.6$} & \multirow{2}{3em}{FKP} & $68\%$ & $31 < f_{\text{NL}} < 89$ & $26 < f_{\text{NL}} < 89$ & $34 < f_{\text{NL}} < 97$ \\ \cmidrule(l){4-7}
         & &  & $95\%$ & $0 < f_{\text{NL}} < 116$ & $-6 < f_{\text{NL}} < 118$ & $6 < f_{\text{NL}} < 129$ \\ \cmidrule(l){3-7}
         & & \multirow{2}{3em}{Optimal} & $68\%$ & $-23 < f_{\text{NL}} < 14$ & $-30 < f_{\text{NL}} < 14$ & $-23 < f_{\text{NL}} < 21$ \\ \cmidrule(l){4-7}
         & & & $95\%$ & $-40 < f_{\text{NL}} < 33$ & $-49 < f_{\text{NL}} < 37$ & $-43 < f_{\text{NL}} < 44$  \\ \midrule \midrule

         \multirow{8}{2em}{NGC} & \multirow{4}{1em}{$1.0$} & \multirow{2}{3em}{FKP} & $68\%$ & $25 < f_{\text{NL}} < 63$ & $24 < f_{\text{NL}} < 65$ & $26 < f_{\text{NL}} < 67$ \\ \cmidrule(l){4-7}
         & & & $95\%$ & $5 < f_{\text{NL}} < 81$ & $3 < f_{\text{NL}} < 85$ & $6 < f_{\text{NL}} < 87$ \\ \cmidrule(l){3-7}
         & & \multirow{2}{3em}{Optimal} & $68\%$ & $-9 < f_{\text{NL}} < 26$ & $-12 < f_{\text{NL}} < 30$ &  $-6 < f_{\text{NL}} < 34$ \\ \cmidrule(l){4-7}
         & & & $95\%$ & $-28 < f_{\text{NL}} < 41$ & $-37 < f_{\text{NL}} < 47$ & $-27 < f_{\text{NL}} < 52$  \\ \cmidrule(l){2-7}
         & \multirow{4}{1em}{$1.6$} & \multirow{2}{3em}{FKP} & $68\%$ & $41 < f_{\text{NL}} < 110$ & $41 < f_{\text{NL}} < 118$ & $49 < f_{\text{NL}} < 125$ \\ \cmidrule(l){4-7}
         & &  & $95\%$ & $10 < f_{\text{NL}} < 145$ & $0 < f_{\text{NL}} < 154$ & $10 < f_{\text{NL}} < 159$ \\ \cmidrule(l){3-7}
         & & \multirow{2}{3em}{Optimal} & $68\%$ & $-41 < f_{\text{NL}} < 16$ & $-48 < f_{\text{NL}} < 20$ & $-39 < f_{\text{NL}} < 28$ \\ \cmidrule(l){4-7}
         & & & $95\%$ & $-75 < f_{\text{NL}} < 40$ & $-89 < f_{\text{NL}} < 47$ & $-80 < f_{\text{NL}} < 54$  \\ \midrule \midrule

         \multirow{8}{2em}{SGC} & \multirow{4}{1em}{$1.0$} & \multirow{2}{3em}{FKP} & $68\%$ & $-10 < f_{\text{NL}} < 42$ & $-17 < f_{\text{NL}} < 37$ & $-15 < f_{\text{NL}} < 41$ \\ \cmidrule(l){4-7}
         & & & $95\%$ & $-35 < f_{\text{NL}} < 67$ & $-41 < f_{\text{NL}} < 66$ & $-36 < f_{\text{NL}} < 72$ \\ \cmidrule(l){3-7}
         & & \multirow{2}{3em}{Optimal} & $68\%$ & $-11 < f_{\text{NL}} < 27$ & $-18 < f_{\text{NL}} < 27$ & $-17 < f_{\text{NL}} < 29$ \\ \cmidrule(l){4-7}
         & & & $95\%$ & $-30 < f_{\text{NL}} < 45$ & $-39 < f_{\text{NL}} < 50$ & $-35 < f_{\text{NL}} < 55$ \\ \cmidrule(l){2-7}
         & \multirow{4}{1em}{$1.6$} & \multirow{2}{3em}{FKP} & $68\%$ & $-20 < f_{\text{NL}} < 72$ & $-30 < f_{\text{NL}} < 67$ & $-22 < f_{\text{NL}} < 78$ \\ \cmidrule(l){4-7}
         & &  & $95\%$ & $-60 < f_{\text{NL}} < 123$ & $-69 < f_{\text{NL}} < 124$ & $-61 < f_{\text{NL}} < 135$ \\ \cmidrule(l){3-7}
         & & \multirow{2}{3em}{Optimal} & $68\%$ & $-22 < f_{\text{NL}} < 31$ & $-36 < f_{\text{NL}} < 27$ & $-28 < f_{\text{NL}} < 36$ \\ \cmidrule(l){4-7}
         & & & $95\%$ & $-43 < f_{\text{NL}} < 63$ & $-57 < f_{\text{NL}} < 68$ & $-51 < f_{\text{NL}} < 74$ \\          
         \bottomrule
    \end{tabular}
    \caption{Summary of the $f_{\text{NL}}$ $68\%$ and $95\%$ constraints of this work. The results for NGC, SGC and the joint analysis are presented, and compared with the eBOSS DR16Q power spectrum constraints \citep{Cagliari:2023mkq}.}
    \label{tab:fnl}
\end{table}

\section{Conclusions}\label{sec:conclusions}

In this work, we presented the local $\fnl$ constraints for the combined tree-level power spectrum and bispectrum analysis of the eBOSS DR16 QSO sample. Assuming that the response to $\fnl$ is proportional to $\delta_c \,(b_1 - p)$ in the first order bias and to $\delta_c \,(b_1 - p) + \left\{ \delta_c \, \left[ b_2 - 8/21 \, (b_1 -1) \right] - b_1 -1  \right\}$ in the second order bias our best constraints at $68\%$ level read as follows
\begin{equation}
    \begin{split}
        -6 < f_{\text{NL}} < 20 \, ,  &\quad \text{for} \, p=1.0 \, , \\
        -23 < f_{\text{NL}} < 14 \, , & \quad \text{for} \, p=1.6 \, .
    \end{split}
\end{equation}
We obtained these bounds with the combination of the power spectrum (optimally weighted) and the bispectrum,  effectively reducing the error bars by $\sim16\%$ with respect to the corresponding optimal $P$-only analysis. This is a novel procedure as in all previous $P+B$ analyses both the power spectrum and bispectrum were FKP weighted. We also tested a compression of the bispectrum, which, on the other hand, did not produce any improvement in comparison to the $P$-only analysis. Additionally, all our results, both improvement- and constraint-wise, are highly consistent with the Fisher forecast we performed before running our analysis. 

Aside from the use of the optimal weights for the power spectrum combined with the bispectrum, an additional novelty of this work is a careful computation of the effective redshift, which can, and often will, be different in the bispectrum compared to the power spectrum, and could lead to biased constraints if not properly taken into account. Additionally, we showed how to include the integral constraint correction to the bispectrum, and found that the radial integral constraint is the largest contribution required in the modeling of the bispectrum. 

There are several possible extensions to this work. First, the implementation of the correct window function convolution in the bispectrum model instead of the approximated version we used here. Second, we still miss an understanding of the optimal weights to extract $\fnl$ from higher-point statistics and the bispectrum in particular. The use of optimal weights for both the power spectrum and bispectrum could be a game changer in the local $\fnl$ measurements from LSS and it is of major importance for the upcoming new data releases by DESI and \textit{Euclid}. We plan to pursue these questions in future works.

\appendix

\acknowledgments

This research benefits from CINECA high performance computing resources and support, under the InDark project, and the High Performance Computing facility of the University of Parma, Italy (HPC.unipr.it), whose support team we thank. 
MSC acknowledges financial support from Fondazione Angelo della Riccia.

\appendix

\section{Two-dimensional posterior distributions}

\begin{figure}
    \centering
    \includegraphics[width=1\linewidth]{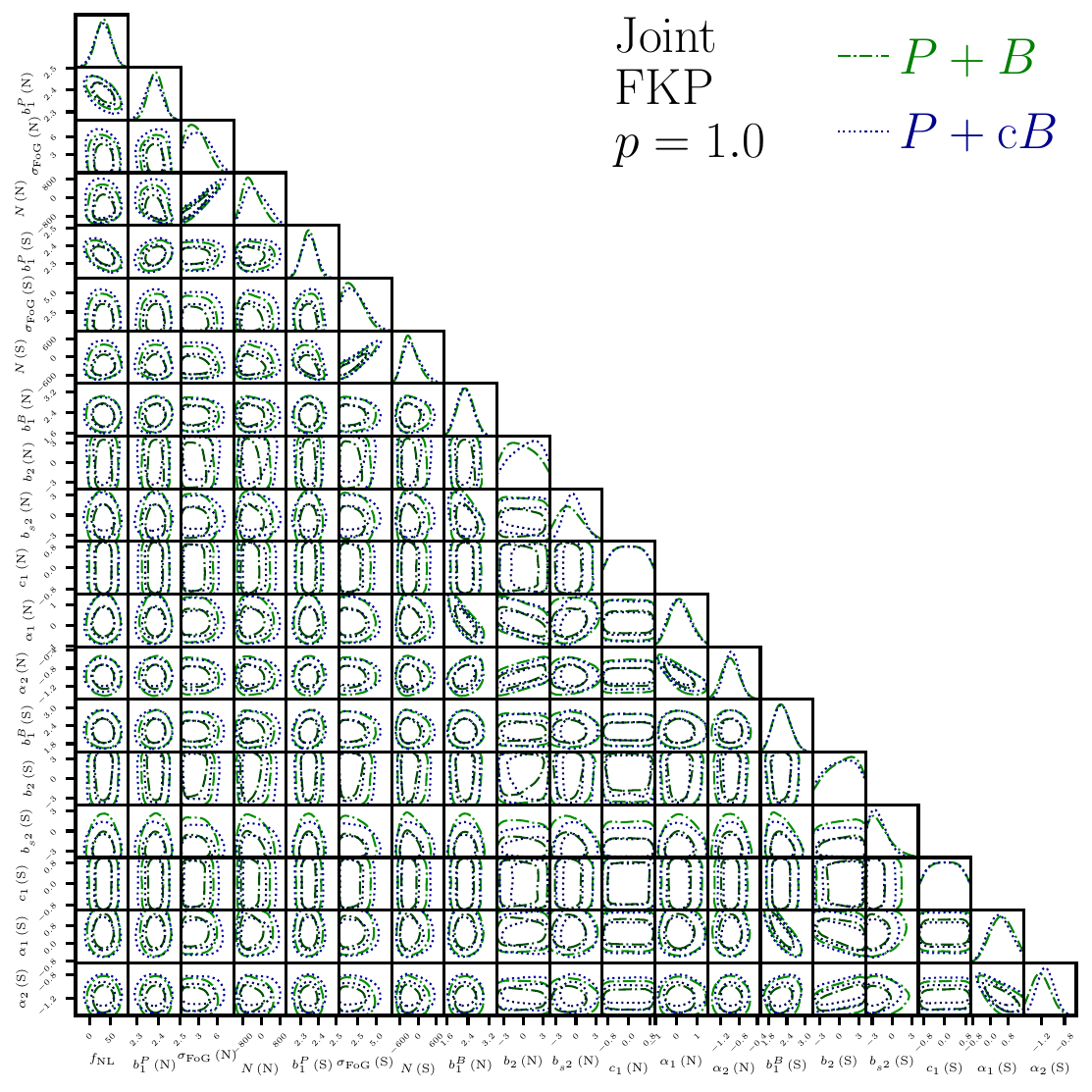}
    \caption{Two-dimensional posterior distribution for $p=1.0$. The dash-dotted green line corresponds to the power spectrum plus bispectrum analysis, while the dotted blue line to the power spectrum plus compressed bispectrum analysis. In both cases, the power spectrum is FKP-weighted. }
    \label{fig:corner-FKP}
\end{figure}

\begin{figure}
    \centering
    \includegraphics[width=1\linewidth]{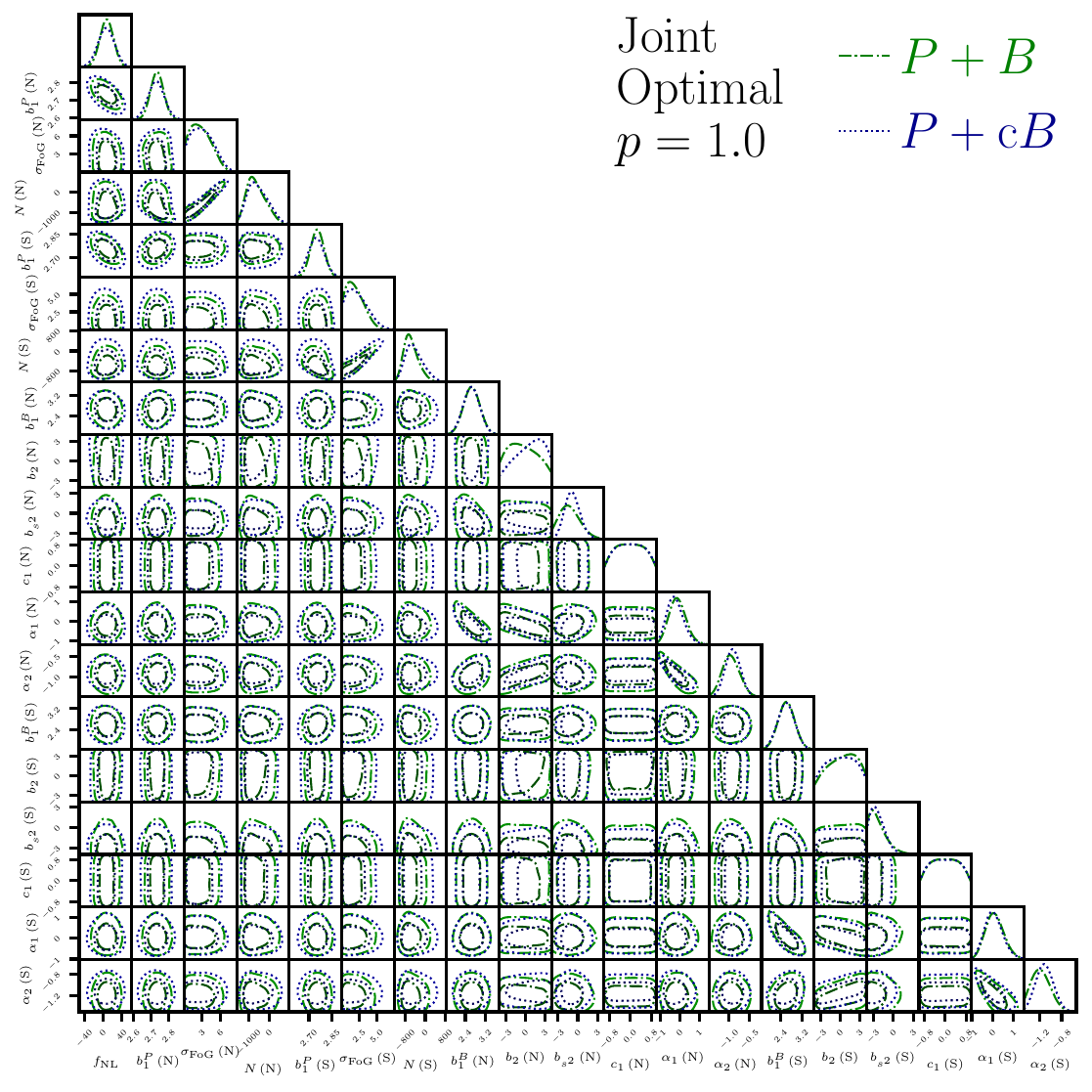}
    \caption{Two-dimensional posterior distribution for $p=1.0$. Same as figure~\ref{fig:corner-FKP}, but the power spectrum is optimally weighted. }
    \label{fig:corner-Opt}
\end{figure}

Figures~\ref{fig:corner-FKP} and \ref{fig:corner-Opt} present the two-dimensional posterior distribution of the $P+B$ (dash-dotted green curves) and $P+{\rm c}B$ (dotted blue curves) analyses. In figure~\ref{fig:corner-FKP} the power spectrum is FKP weighted, while in figure~\ref{fig:corner-Opt} it is optimally weighted. 

In both cases, the nuisance parameter distributions of the two analyses are consistent. We note that the $b_2$ and $c_1$ posterior distributions are prior-dominated and we have no constraining power over these parameters. For testing purposes, we also run our analysis setting larger priors on these two parameters ($b_2 \in [-20,20]$ and $c_1 \in [-100,100]$), which are consistent with the Fisher forecast. The use of these priors did not produce any change in the $\fnl$ posterior distribution and, consistently with the Fisher analysis, we were able to constrain $b_2$ and $c_1$. However, due to the larger prior space, the chain computational time greatly increased. Therefore, to reduce the computational expenses, we opted for tighter prior distribution for these parameters.

Finally, we note that in both cases, the distribution of the first-order biases for $P$ and $B$ are consistent with the use of a larger $z_{\rm eff}$ for $P$ over $B$.

\section{Test of the bispectrum-window convolution approximation on the mocks}
\label{app:window_mocks}

\begin{figure*}
\centering
\includegraphics[width=\columnwidth]{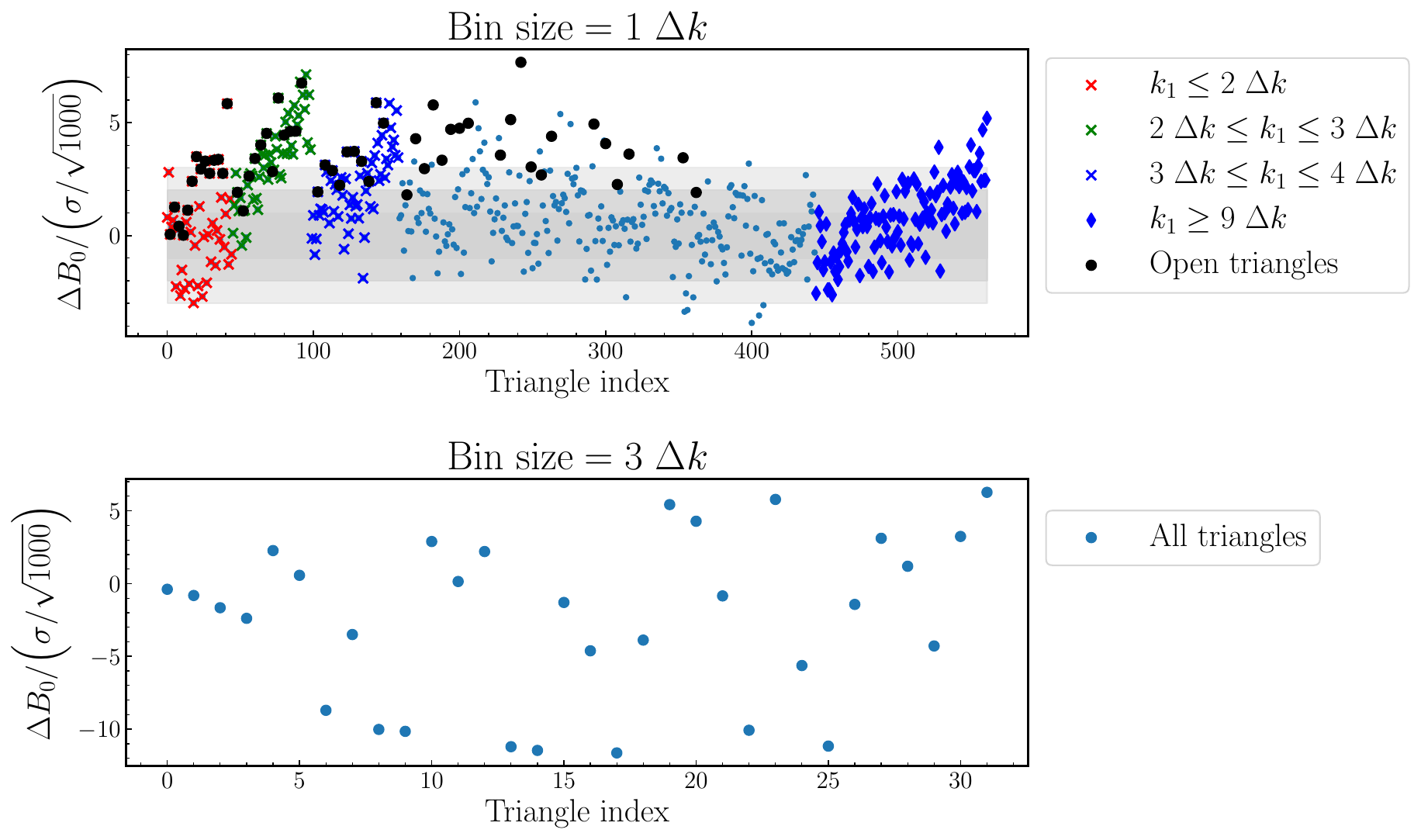}
   \caption{The difference between the best-fit model (with  $f_\mathrm{NL} = 0$) and the mean of the NGC EZmock is compared against the standard deviation of the mean of the mocks, $\sigma_\mathrm{mean} = \sigma/\sqrt{1000}$. This is based on a cumulative volume equal to 1000 times larger than the effective volume of the NGC, where $\sigma$ represents the standard deviation of the data. A constant shot noise has been subtracted from the data. The triangles are ordered so that $k_3$ is varied first, followed by $k_2$, and then $k_1$, with the constraint $ k_1 \leq k_2 \leq k_3$. The upper panel compares data with bin size $\Delta k$, while the bottom panel shows the same comparison for $3 \, \Delta k$.}
\label{fig:window_check_mocks}
\end{figure*}

We conduct a stringent test of the approximate bispectrum window-convolution by comparing the difference between the best-fit model (with $\fnl=0$) and the mean of the mocks to the standard deviation of the mock means, $\sigma_\mathrm{mean}$. Notably, $\sigma_\mathrm{mean}$ corresponds to a cumulative volume 1000 times larger than the data's effective volume, yielding $\sigma_\mathrm{mean} \approx \sigma/32$, where $\sigma$ is the data's standard deviation, highlighting the test's rigor. To better explore the dependence on triangle shapes, we perform this test on the bispectrum measurement using the original $\Delta k$ binning. As shown in the upper panel of figure~\ref{fig:window_check_mocks}, the model generally agrees with the data within the $3\,\sigma_\mathrm{mean}$ region, though deviations appear mainly for squeezed triangles. The bottom panel presents the same comparison for data binned in $3\Delta k$, where the theory aligns with the data within approximately $12 \, \sigma/\sqrt{1000}$. 

\bibliographystyle{JHEP}
\bibliography{main}

\end{document}